\PassOptionsToPackage{pdftex,pdfa,pdfstartview=,bookmarks=true}{hyperref}

\documentclass[sigconf, nonacm]{acmart}

\usepackage{xspace}
\usepackage{paralist}
\usepackage{array}
\usepackage{arydshln} 
\usepackage{multirow}
\usepackage{numprint}
\usepackage{ifthen}
\usepackage{algorithm}
\usepackage{algorithmic}
\usepackage{enumitem}
\usepackage{balance}
\usepackage{booktabs}
\usepackage{makecell}
\usepackage[a-2b,mathxmp]{pdfx}[2018/12/22]
\usepackage{colorprofiles}

\newcommand{\true}{true}
\newcommand{\false}{false}
\newcommand{\PaperLongVersion}{\true}

\newcommand{\PaperLong}[1]
{
  \ifthenelse{\equal{\PaperLongVersion}{\true}}{#1}{\empty}
}
\newcommand{\PaperShort}[1]
{
  \ifthenelse{\equal{\PaperLongVersion}{\false}}{#1}{\empty}
}

\newcommand{\snowman}{Snowman\xspace}
\newcommand{\snowmans}{Snowman's\xspace}
\newcommand{\Snowman}{\snowman}
\newcommand{\Snowmans}{\snowmans}
\newcommand{\frost}{Frost\xspace}
\newcommand{\Frost}{\frost}
\newcommand{\frosts}{\frost's}

\newcommand{\benchmarkPlatform}{benchmark platform\xspace}

\newcommand{\aBenchmarkPlatform}{a benchmark platform\xspace}
\newcommand{\benchmark}{benchmark\xspace}
\newcommand{\benchmarks}{benchmarks\xspace}
\newcommand{\Benchmarks}{Benchmarks\xspace}

\newcommand{\TwoElemSubsets}[1]{[#1]^2}
\newcommand{\fMeasure}{f1 score\xspace}
\newcommand{\fStarMeasure}{f* score\xspace}

\newcommand*\colvec[3][]{
    \begin{pmatrix}\ifx\relax#1\relax\else#1\\\fi#2\\#3\end{pmatrix}
}

\newcommand{\backend}{back-end\xspace}
\newcommand{\Backend}{Back-end\xspace}
\newcommand{\frontend}{front-end\xspace}
\newcommand{\Frontend}{Front-end\xspace}

\newcommand{\captionWithDescription}[2]{\caption[#1]{\textbf{#1.} \normalfont{#2}}}
\newcommand{\itemWithTitle}[2]{\item \textbf{#1}: #2}

\newcommand\vldbdoi{10.14778/3554821.3554823}
\newcommand\vldbpages{3292 - 3305}
\newcommand\vldbvolume{15}
\newcommand\vldbissue{12}
\newcommand\vldbyear{2022}
\newcommand\vldbauthors{\authors}
\newcommand\vldbtitle{\shorttitle} 
\newcommand\vldbavailabilityurl{https://github.com/HPI-Information-Systems/snowman}
\newcommand\vldbpagestyle{plain} 

\begin{document}

\title{\frost: A Platform for Benchmarking and Exploring Data Matching Results}

\author{Martin Graf}
\email{martin.graf@hpi-alumni.de}
\affiliation{%
  \institution{Hasso Plattner Institute}
  \country{University of Potsdam, Germany}
}
\author{Lukas Laskowski}
\email{lukas.laskowski@student.hpi.de}
\affiliation{%
  \institution{Hasso Plattner Institute}
  \country{University of Potsdam, Germany}
}

\author{Florian Papsdorf}
\email{florian.papsdorf@student.hpi.de}
\affiliation{%
  \institution{Hasso Plattner Institute}
  \country{University of Potsdam, Germany}
}

\author{Florian Sold}
\email{florian.sold@student.hpi.de}
\affiliation{%
  \institution{Hasso Plattner Institute}
  \country{University of Potsdam, Germany}
}

\author{Roland Gremmelspacher}
\email{roland.gremmelspacher@sap.com}
\affiliation{
  \institution{SAP SE}
  \city{Walldorf}
  \country{Germany}
}

\author{Felix Naumann}
\email{felix.naumann@hpi.de}
\affiliation{%
  \institution{Hasso Plattner Institute}
  \country{University of Potsdam, Germany}
}

\author{Fabian Panse}
\email{fabian.panse@uni-hamburg.de}
\affiliation{%
  \institution{University of Hamburg, Germany}
}

\begin{abstract}
``Bad'' data has a direct impact on 88\% of companies, with the average company losing 12\% of its revenue due to it.
Duplicates -- multiple but different representations of the same real-world entities -- are among the main reasons for poor data quality, so finding and configuring the right deduplication solution is essential.
Existing data matching benchmarks focus on the quality of matching results and neglect other important factors, such as business requirements.
Additionally, they often do not support the exploration of data matching results.

To address this gap between the mere counting of record pairs vs.\ a comprehensive means to evaluate data matching solutions, we present the \frost platform.
It combines existing \benchmarks, established quality metrics, cost and effort metrics, and exploration techniques,
making it the first platform to allow systematic exploration to understand matching results.
\frost is implemented and published in the open-source application \snowman, which includes the visual exploration of matching results, as shown in Figure~\ref{fig:snowmanIntersection}.
\end{abstract}

\maketitle

\pagestyle{\vldbpagestyle}
\begingroup\small\noindent\raggedright\textbf{PVLDB Reference Format:}\\
\vldbauthors. \vldbtitle. PVLDB, \vldbvolume(\vldbissue): \vldbpages, \vldbyear.\\
\href{https://doi.org/\vldbdoi}{doi:\vldbdoi}
\endgroup
\begingroup
\renewcommand\thefootnote{}\footnote{\noindent
This work is licensed under the Creative Commons BY-NC-ND 4.0 International License. Visit \url{https://creativecommons.org/licenses/by-nc-nd/4.0/} to view a copy of this license. For any use beyond those covered by this license, obtain permission by emailing \href{mailto:info@vldb.org}{info@vldb.org}. Copyright is held by the owner/author(s). Publication rights licensed to the VLDB Endowment. \\
\raggedright Proceedings of the VLDB Endowment, Vol. \vldbvolume, No. \vldbissue\ %
ISSN 2150-8097. \\
\href{https://doi.org/\vldbdoi}{doi:\vldbdoi} \\
}\addtocounter{footnote}{-1}\endgroup

\ifdefempty{\vldbavailabilityurl}{}{
\vspace{.3cm}
\begingroup\small\noindent\raggedright\textbf{PVLDB Artifact Availability:}\\
The source code, data, and/or other artifacts have been made available at \url{\vldbavailabilityurl}.
\endgroup
}

\section{Data Matching}
\label{sec: Introduction}

Businesses and organizations rely heavily on structured data in databases.
These databases often contain errors, such as outdated values, typos, or missing information, leading to large costs and non-monetary damage~\cite{davis2014TheCostOfBadData}.
One prominent aspect of inaccurate data is (fuzzy) duplicates -- the presence of multiple but different records representing the same real-world entity.
Beyond sloppy data entry, duplicates emerge in further situations, in particular when integrating data from multiple sources.
\begin{figure}[t]
\centering
\fbox{\includegraphics[trim={0 0 2.5cm 0}, clip, width=\columnwidth]{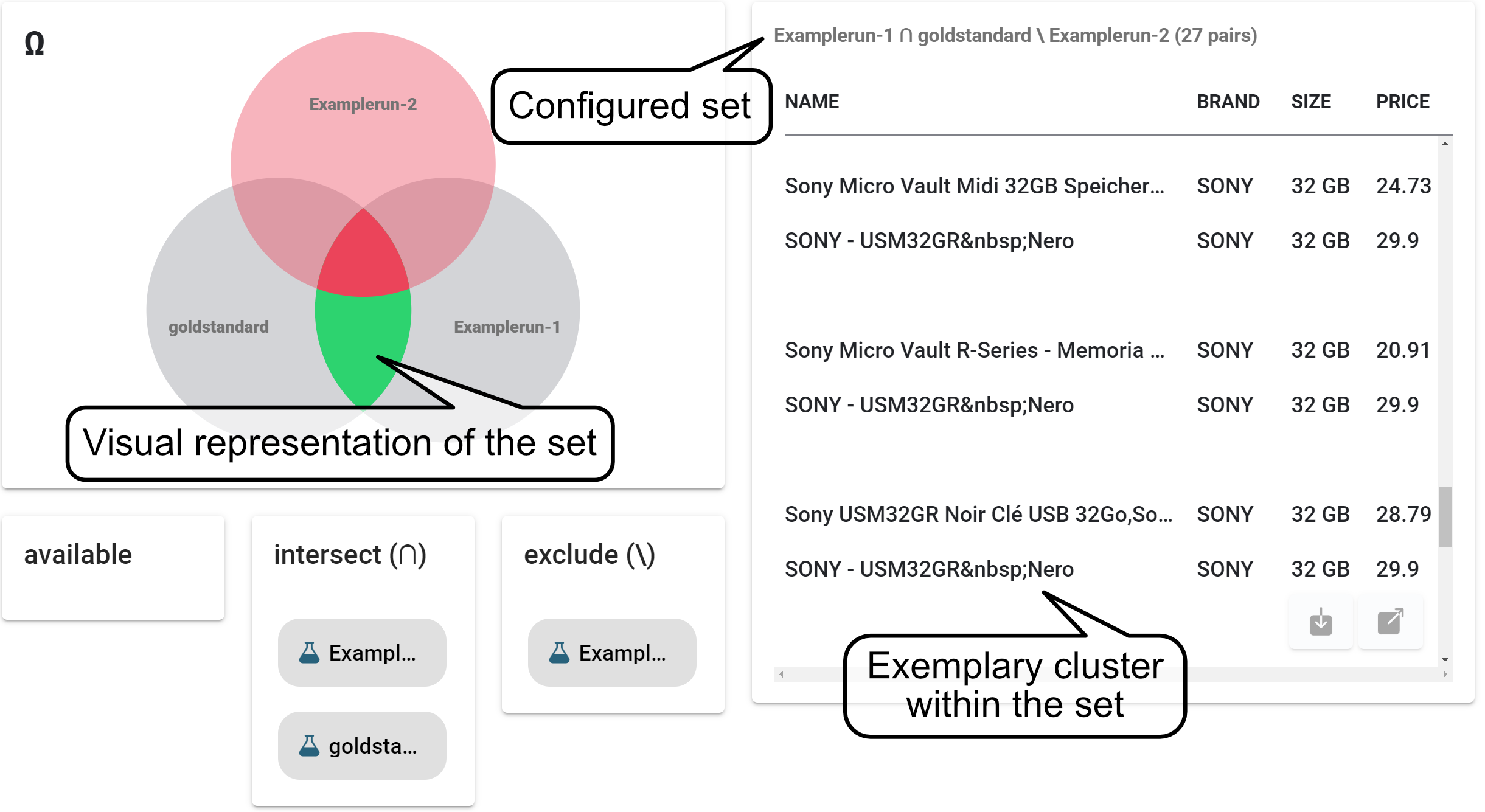}}
\captionWithDescription{Exploring data matching results in \snowman}{This figure shows the ground truth matches that Examplerun-1 found and Examplerun-2 did not find.}
\label{fig:snowmanIntersection}
\end{figure}
To address the issue of duplicates, various commercial and research systems to detect such duplicates have been developed~\cite{ChristophidesEP21,Christen2012DataMatching}.

Systems detecting duplicates are generally referred to as (data) matching solutions, deduplication solutions, or entity resolution systems.
They can be broadly categorized into two groups. 
Rule-based solutions are configured by hand-crafted matching rules to detect when a pair of records is a duplicate.
An example rule in the context of a customer dataset could state that a high similarity of the surname is an indicator for duplicates, but a high similarity of customer IDs is not.
Supervised machine learning models, on the other hand, are trained by domain experts who label example pairs from the dataset as \emph{duplicate} or \emph{non-duplicate}.

\subsection{A Data Matching Benchmark Platform}
\label{ssec: Contributions}

To find the best matching solution for a specific use case and to help configure it optimally, different benchmarks for comparing matching results have been developed.
Typical data matching benchmarks consist of a dataset, a ground truth annotation, and sometimes information on how to evaluate the performance.
To answer the question of which matching solution is best, all competing matching solutions are run against the dataset.
Then, the results are compared to the ground truth annotation.
Finally, performance metrics scores, such as precision, recall, and \fMeasure are determined for each matching solution and contrasted.

While many such benchmarks exist (see Section~\ref{sec: Related Work}), the community is lacking a platform to easily (and interactively) compare the results of running against multiple such benchmarks, or to easily compare multiple systems or multiple system configurations against the same benchmark.

Almost all existing matching solution evaluation techniques focus on the quality of matching solution results.
However, other aspects, such as business factors, also affect their usefulness in real-world deployment scenarios.
Traditional metrics often provide only a quantitative overview over the performance of matching solutions.
Qualitative analyses, such as behavioral analytics for matching solutions, are scarce in the matching context -- despite having a high relevance for common use cases, such as fine-tuning a matching solution.
To address these issues, we make the following contributions to the field of data matching benchmarks, noting that we do not propose yet another benchmark or further benchmark datasets, but rather a platform to systematically explore and analyze data matching results, thus enabling (business) users to more easily assess and compare their results for any given dataset.

Our extensible data matching benchmarking platform \textbf{\frost} offers both traditional quantitative quality metrics, but adds effort measurements by aggregating relevant business factors, for example purchase costs and deployment type.
It includes techniques to systematically explore and compare matching results, allowing qualitative inter-system and intra-system inspection. 
The combination of traditional quality metrics, soft key-performance-indicators (KPIs) and exploration techniques allows deep evaluations for the industrial context.
We implemented the majority of \frost in the open-source application \href{https://github.com/HPI-Information-Systems/snowman}{\textbf{\snowman}}
and prove its practical relevance within the industrial context by demonstrating the usability of the platform and efficacy of the resulting evaluation insights.
\begin{itemize}[leftmargin=.5cm]
    \item{To suit the enterprise context, \snowman is meant to operate in, we chose and optimized evaluation algorithms for a \textit{completely portable, self-bundled application stack}, which does not require privileged permissions for installation or execution that might be difficult to acquire.}
    \item{The extensibility of \snowman with \textit{additional data import formats as well as new evaluation techniques} is an intrinsic part of its modular architecture.}
    \item{Apart from traditional evaluation metrics, \snowman supports \emph{exploration techniques that do not require a ground truth} for industry use cases with datasets lacking a ground truth.}
    \item{In a variety of projects at SAP, we were able to observe that \snowmans user interface fosters \textit{engagement with data matching stakeholders beyond IT}. Therefore, we consider it to be the reference implementation of the ideas presented in \frost and a starting point for open-source collaboration.}
    \item{Snowman does not execute the matching solutions itself, but takes their results as input, which it then matches against the ground truth. 
    This allows Snowman to be used quickly in many different environments with low resource requirements.
    }
\end{itemize}

\subsection{Formal Matching Process}
\label{ssec: Formal Matching Process}

In this section, we define the formal matching process, and  introduce abbreviations and variables that we use throughout the paper.

A dataset $D$ is a collection of records that may contain duplicates.
A record pair is a set of two records $\{r_1, r_2\} \subseteq D$.
We denote the set of all record pairs in $D$ as $\TwoElemSubsets{D} =\{ \, A \subseteq D \: | \: |A| = 2 \, \}$.
A matching solution $M$ is a function which takes a dataset $D$ as input and outputs a disjoint clustering $\{C_{\mathit{1}}, C_{\mathit{2}}, ...\}$ of $D$.
We call the output of a matching solution an \emph{experiment}.

All pairs \emph{within} a cluster $C_{\mathit{i}}$ are predicted to be duplicates by the matching solution and called \emph{matches}.
All other pairs in $\TwoElemSubsets{D}$ are predicted to be no duplicates and called \emph{non-matches}.
Accordingly, a different representation for the clustering is a set of all matches $E \subseteq \TwoElemSubsets{D}$.
$E$ can be seen as a graph with a node for every record $r \in D$ and edges between all record pairs $\{r_1, r_2\} \in E$ (also called \emph{identity link network}~\cite{Idrissou2020NoGoldstandard}).
Because $E$ represents a clustering of $D$, the graph is transitively closed, ensuring that if $r_1$ and $r_2$ are matches and $r_2$ and $r_3$ are matches, $r_1$ and $r_3$ are considered to be matches, too.
Nevertheless, some real-world matching solutions output subsets of $\TwoElemSubsets{D}$ that are not transitively closed.
Although the closure of such a result set could easily be created, this step often introduces many false positives~\cite{Hassanzadeh2009ClusterAlgorithms, Draisbach2019PairsToClusters}.
Instead, a clustering algorithm specific to the use case can be applied~\cite{Hassanzadeh2009ClusterAlgorithms,Draisbach2019PairsToClusters}.

While it is most common to evaluate the performance of matching solutions only via their final results, typical matching solutions consist of multiple steps~\cite{PanseNaumann2021EvaluationofDedupAlgos}.
Measuring the performance between these steps, as supported by \frost, can provide useful insights for tweaking specific parts of the matching solution and helps to find bottlenecks of matching performance. 
A data matching pipeline typically consists of the following steps:
\begin{enumerate}[leftmargin=.5cm]
\itemWithTitle{Data preparation}{Segment, standardize, clean, and enrich the original dataset~\cite{koumarelas2020DataPreparation}.}

\itemWithTitle{Candidate generation}{Create subset of candidate pairs that contains as many true duplicates as possible, for instance using blocking or windowing~\cite{Christen2012IndexingTechniques, PapadakisSTP20}.}

\itemWithTitle{Similarity-based attribute value matching}{Compute similarities between the records' attribute values for each candidate pair~\cite{Ives2012DataIntegration, Christen2012DataMatching}.}

\itemWithTitle{Decision model / classification}{Given the similarities for each candidate pair, decide which candidate pairs are probably duplicates~\cite{Ives2012DataIntegration, Christen2012DataMatching}.
Typically, this step produces a final similarity or confidence score for each candidate pair.
A pair is matched if its score is higher than a specific threshold.
We use the term \emph{similarity} to refer to both similarity and confidence.}

\itemWithTitle{Duplicate clustering}{Given the set of high probability duplicate pairs, cluster the original dataset into disjoint sets of duplicates~\cite{Hassanzadeh2009ClusterAlgorithms, Draisbach2019PairsToClusters}.}

\itemWithTitle{Duplicate merging / record fusion}{Merge the clusters of duplicates into single records~\cite{Bleiholder2009DataFusion, Deng2017GoldenRecord, heidari2020recordFusion}}
\end{enumerate}

In the following sections, we first discuss related work (Section~\ref{sec: Related Work}).
Then, we discuss different means for benchmarking, including traditional quality measures and soft KPIs (Section~\ref{sec: Benchmarking Data Matching Solutions}).
We introduce different evaluation techniques that allow for qualitative insights about matching solutions in Section~\ref{sec: Exploring Data Matching Results}.
Finally, we showcase \snowman, our implementation of \frost, and present a short study on the impact of effort measurements in Section~\ref{sec: Reference Implementation}.\PaperShort{A long version of this paper with further architectural details is available at~\cite{frost2022arxiv}.}

\section{Related Work}
\label{sec: Related Work}

We outline existing work on benchmarking and exploring matching solutions.
First, we discuss existing \benchmarks, then approaches that yield insights similar to our exploration techniques, and lastly other benchmark platforms. Please note that reviewing related work on actual matching solutions, such as JedAI~\cite{Papadakis20} or Magellan~\cite{Doan20}, is beyond the scope of this paper, and we refer to corresponding surveys~\cite{Christen2012DataMatching,Papadakis21}.

\subsection{Data Matching Benchmarks}
\label{ssec: Benchmark Datasets (related)}

\Benchmarks evaluate the performance of matching solutions, laying the foundation for major parts of our \benchmarkPlatform \frost.
We discuss benchmark datasets in more detail in Section~\ref{ssec: Benchmark Datasets}.
A list of prominent \benchmarks, generators, and polluters is collected in~\cite{PanseNaumann2021EvaluationofDedupAlgos}.
Below, we discuss two recent \benchmarks that are especially relevant to \frost.

The \textit{Semantic Publishing Instance Matching Benchmark} (SPIMBench) consists of a data generator capable of value, structural, and logical transformations producing a \emph{weighted gold standard} and a set of metrics for evaluating matching performance~\cite{Saveta2014SPIMBenchAS}.
The weighted gold standard includes a history of which transformations were applied to generated duplicates to allow a detailed error analysis.
While SPIMBench is helpful to optimize duplicate detection within RDF datasets, it cannot be used with relational data that is common in our industrial context.

Crescenzi et al.\ propose the flexible schema matching and deduplication \benchmark Alaska~\cite{crescenzi2021alaska}.
The authors profiled their datasets with traditional profiling metrics and three new metrics for measuring heterogeneity, namely \emph{attribute sparsity}, \emph{source similarity}, and \emph{vocabulary size}.
In work from 2020 Primpeli and Bizer focused solely on profiling benchmark datasets and grouped 21 \benchmarks according to five profiling metrics, namely \emph{schema complexity}, \emph{textuality}, \emph{sparsity}, \emph{development set size}, and \emph{corner cases}~\cite{Primpeli2020ProfilingMatchingTasks}.
Such profiling metrics allow for better comparability of quality metrics from different \benchmarks because the profiled factors can be considered.
Moreover, profiling metrics that measure the difficulty or heterogeneity of datasets are a crucial step towards finding representative datasets for a given matching task when no ground truth annotations exist.

\frost supports a wide range of profiling metrics for measuring how similar a benchmark dataset and a real-world dataset are (see Section~\ref{sssec: Finding a Representative Benchmark Dataset}).
Additionally, we use the notion of attribute sparsity proposed by Crescenzi et al.\ for classifying errors of matching solutions (see Section~\ref{ssec: Error Analysis}).

\subsection{Exploration Opportunities}
\label{ssec: Exploration Opportunities (related)}

There has been surprisingly little work on techniques to systematically explore and understand matching results.
SIMG-VIZ interactively visualizes large similarity graphs and entity resolution clusters~\cite{Rostami2018InteractiveVO}, helping users to detect errors in the duplicate clustering stage.
This is useful for improving the clustering algorithm and can give an overview on the matching result.
Yet, only a limited number of possible errors are highlighted and large graphs easily overwhelm users.
To counteract these problems, we propose techniques that help users to detect errors within the decision model.
Specifically, we reduce the amount of information presented to the user by filtering out irrelevant data, sorting it by interestingness (Section~\ref{sec: Exploring Data Matching Results}), and enriching it with useful information about the error.

NADEEF/ER introduces additional investigation techniques for the rule-based approach NADEEF~\cite{DBLP:conf/sigmod/ElmagarmidIOQ0Y14}.
NADEEF/ER offers users a complete suite for rule-based entity matching, including an exploration dashboard to analyze matching results, for example the influence of each individual rule on the result.
Compared to NADEEF/ER, \frost uses a more generic approach to evaluate matching results, as it supports a broad variety of matching solutions.

Matching solutions utilizing active learning, such as proposed by Sarawagi and Bhamidipaty~\cite{sarawagi2002ActiveLearning}, try to minimize review cost by asking human annotators only about uncertain matching decision; they are shown only uncertain matching decisions, and thus, they can understand weaknesses of the matching solution.
Qian et al.\ introduced SystemER~\cite{QianSystemER} as an active-learning-based entity resolution pipeline that uses solely rules comprehensible to humans, thus explaining individual matching decisions. 
\subsection{Benchmark Platforms}
\label{ssec: Benchmark Platforms}
In~\cite{Weis2006ADD}, the authors described an early benchmark platform for XML data, measuring effectiveness (matching quality) and efficiency (runtime).
\frost integrates both effectiveness and efficiency measurements, but is not limited to them.

A new measurement dimension, \emph{effort}, was proposed in the benchmark platform FEVER~\cite{Koepcke2009FEVER}.
Next to quality metrics, such as precision and recall, FEVER allows measuring the effort to configure a matching solution run by specifying labeling and parametrization effort.
These KPIs can be compared in effort-metric diagrams, answering questions such as ``How much effort is needed to reach 80\% precision?''
\frost builds on this idea by integrating business requirements to support the decision-making process of selecting a matching solution. For instance, it allows the comparison of matching solutions based on context-sensitive effort measurements, but also on further KPIs, such as deployment type and costs.

In later work, the authors used FEVER to evaluate different matching solutions.
They found that ``some challenging resolution tasks such as matching product entities from online shops are not sufficiently solved with conventional approaches based on the similarity of attribute values''~\cite{Koepcke2010UsingFEVER}.
This insight emphasizes the need for a comprehensive \benchmarkPlatform and the ability to systematically explore matching results.

Another benchmark platform, GERBIL, is based on the BAT framework~\cite{cornolti2013framework} and provides 46 datasets, 20 matching solutions and eight tasks~\cite{roder2018gerbil}. New matching solutions can be integrated by conforming to a REST API. Afterwards, they can be evaluated with the available datasets and tasks. The importance of such a platform is outlined by the fact that the GERBIL community already carried out more than 24,000 evaluations with GERBIL. While GERBIL is useful for evaluating quality metrics in different scenarios, it does not provide a way to compare soft KPIs or to explore matching results. Thus, we see it as very useful for finding the best among a selection of matching solutions. \Frost also provides a selection of these quality metrics. However, it integrates evaluations allowing developers to gradually improve their matching solutions, such as metric-metric diagrams (see section~\ref{sssec: Metric-Metric Diagrams}), as well.

Some data matching execution frameworks also work as (partial) benchmark platforms.
\emph{DuDe} is a modular duplicate detection toolkit~\cite{Draisbach2010DuDeTD}, consisting of six components to facilitate the entire matching process.
One of those components, the postprocessor, can evaluate the performance of the matching solution in each run.
For this purpose, metrics, such as precision and recall, are calculated.
This reduces the feedback loop between running a matching solution and interpreting performance results, and allows comparing different experiments performed with DuDe.
On the other hand, DuDe does not support general comparability between matching solutions, because many matching solutions use other matching frameworks or do not use a framework at all.
A newer approach to this concept is the weakly supervised Panda platform~\cite{WuSLCH21}, which uses user-created labelling functions to solve a given matching task. It allows for constant feedback on the performance of a certain labeling function and also offers debugging tools. Nevertheless, its use-cases are rather limited, as it also does not allow general comparability between arbitrary matching solutions.

\section{Benchmarking Matching Solutions}
\label{sec: Benchmarking Data Matching Solutions}

\frost is a platform that supports users in evaluating their matching solutions using arbitrary data matching \benchmarks.
A data matching \benchmark typically consists of
\begin{itemize}[leftmargin=.5cm]
\item{ One or more dirty \emph{datasets} containing duplicates.
These duplicates can be within, but also between, the individual datasets (intra-source vs.\ inter-source duplicates).}

\item{A \emph{gold standard} 
modeling the ground truth,  i.e., the correct duplicate relationships between the given data records}.

\item{A set of \emph{quality metrics} to evaluate the given matching solutions.
These can be metrics that compare these solutions' results with the gold standard,
such as recall or precision~\cite{Menestrina2010EvaluatingResults},
but also metrics that measure some inherent properties of these results,
e.g., the number of pairs that are missing to transitively close the set of discovered matches
or some soft KPIs, such as the effort that is needed to compute them.}
\end{itemize}

\subsection{Benchmark Datasets}
\label{ssec: Benchmark Datasets}

A good \benchmark should meet several conditions.
\begin{inparaenum}[(i)]
\item
Most importantly, its ground truth annotation should be as correct and complete as possible (see Section~\ref{sssec: Gold Standards}).
\item
Second, to generalize well, 
its data and error patterns should be real or at least realistic (see Section~\ref{sssec: Reference Datasets}).
\item
Third, the dataset should contain some so-called \emph{corner cases}~\cite{Primpeli2020ProfilingMatchingTasks} 
to push matching solutions to their limits and reveal their quality differences.
\item
Finally, the dataset must be compatible with the objectives of the evaluation.
For example, evaluating a matching solution focused on scalability
requires a large dataset with millions of records,
while the evaluation of a clustering algorithm 
requires a dataset with duplicate-clusters of various sizes.
\end{inparaenum}

\subsubsection{Gold Standards}
\label{sssec: Gold Standards}
To measure the correctness of an experiment, we need a reference solution against which we can compare the result of the experiment.
This solution is also called \emph{gold standard} or \emph{ground truth} and should accurately reflect the true state of the real-world scenario as defined by the use case (e.g., matching by household vs.\ by person).

The truth about the correct duplicate relationships between the records of a dataset $D$ can be captured in different ways.
The most common approach is to store a list of all pairs of duplicate records (or their IDs respectively) in a separate file.
The gold standard, however, typically represents complete knowledge about the correct duplicate relationships and thus corresponds to a final matching solution~\cite{Menestrina2010EvaluatingResults}, i.e., it is a clustering of $D$ where every record belongs to exactly one cluster.
Thus, the gold standard can also be modeled within the actual dataset by adding an extra attribute that associates each record with its corresponding cluster.
\frost supports both formats, making importing new gold standards as easy as possible.

\subsubsection{Reference Datasets}
\label{sssec: Reference Datasets}
Many users who need a \benchmarkPlatform have their own use cases with their own datasets.
Since the true duplicate relationships within these datasets are usually unknown (this is, after all, the reason matching solutions are applied), the performance of a matching solution cannot be evaluated on the whole dataset of the use case itself.
Instead, the evaluation is frequently performed on a small subset of the dataset or on a similar reference dataset (see Section~\ref{sssec: Finding a Representative Benchmark Dataset}).
Sometimes, as is also the case within SAP, manually annotated datasets from previous cleaning processes are available and can be reused.

Reference datasets can originate from the real world or can be artificially created~\cite{PanseNaumann2021EvaluationofDedupAlgos}.
In real-world datasets, the true duplicate relationships need to be labeled by domain experts.
The data matching community has compiled several such datasets over the past few decades, which are publicly available
via various sources, such as the Magellan Data Repository~\cite{dasXXXXMagellanData}.
The artificial creation of test data can be automated.
Examples of such test data generators are
TDGen~\cite{Bachteler12}, GeCo~\cite{ChristenV13}, LANCE~\cite{SavetaDFFN15},  BART~\cite{ArocenaGMMPS15}, or EMBench++~\cite{IoannouV19}. 

Our reference implementation \snowman already includes a number of popular benchmark datasets, such as Cora and CDDB~\cite{Draisbach2010DuDeTD}.
However, to allow easy use of any dataset (whether real-world or artificially created), it also supports an easy creation of custom importers (see Section~\ref{ssec: Snowman Features}).

\subsubsection{Finding a Representative Benchmark Dataset}
\label{sssec: Finding a Representative Benchmark Dataset}

Researchers usually want to test their newly developed matching solutions under different conditions, and therefore like to use benchmark datasets that differ in their characteristics. 
To achieve this, they either create these artificially with the help of generators or make use of the numerous datasets provided by the community.
In contrast, practitioners usually do not develop new matching solutions,
but must use an existing one to detect all duplicates within their use case-specific datasets.
Thus, while researchers want to evaluate a particular solution on different datasets, practitioners aim to evaluate different solutions on a particular dataset.
Therefore, practitioners cannot just take any benchmark dataset for their evaluation, but strive to find one that is similar to their use case dataset 
so that they can estimate the performance of different matching solutions on the latter by evaluating them on the first.
Logically, the performance estimated in this way is only meaningful if both datasets pose similar challenges to the matching solutions.
This makes finding a suitable benchmark dataset difficult.

To assist users in this search process, \frost includes a list of features impacting matching difficulty
and provides decision matrices to compare a given use case dataset with several benchmark datasets based on these features.
It remains to the experts to determine how important the individual features are for their use case, 
and to select the benchmark dataset that they think is best suited for their evaluation goals.
In addition to the features proposed by Crescenzi et al.~\cite{crescenzi2021alaska} 
and Primpeli et al.~\cite{Primpeli2020ProfilingMatchingTasks},
such as sparsity, textuality, and schema complexity, additional useful features are:
\begin{itemize}[leftmargin=.5cm]
    \itemWithTitle{Domain}{The domain of both datasets should match or be closely related.}

    \itemWithTitle{Record count}{Draisbach and Naumann showed that dataset size has influence on the optimal similarity threshold~\cite{Draisbach2013ChoosingThresholds}.
    Thus, using a benchmark dataset with similar size compared to the use case dataset may yield more representative results.}

    \itemWithTitle{Number and size of duplicate clusters}{
    The amount and size of duplicate clusters in the ground truth annotation of the benchmark dataset should closely resemble that of the use case dataset.
    Because the ground truth annotation for the use case dataset is unknown, these numbers have to be estimated.
    Heise et al.\ developed a method for this estimation~\cite{2014HeiseEstimatingNumberOfDuplicates}.}

    \itemWithTitle{Matching solution}{The matching solution itself may provide valuable insights into how similar 
    both datasets are. Relevant features include 
\begin{inparaenum}[(i)]
    \item metrics for approximating quality without requiring a ground truth annotation (see Section~\ref{sssec: Evaluating Quality Without Ground Truth}), 
	\item the similarity of the clusterings of the matching solution on use case and benchmark dataset, and 
	\item the number of pairs from the transitive closure that are missing in the solution's classification results on both datasets.
	\end{inparaenum}
    Note that some of these metrics require normalization if certain properties of the datasets, such as record count, do not match.}

\itemWithTitle{Vocabulary similarity}{Vocabulary similarity $\textit{VS}$ quantifies the similarity of the vocabularies of two datasets.
    Similar vocabularies might cause similar behavior of the matching solution.
    We calculate this similarity using the Jaccard coefficient: 
    \[
    \textit{VS}(D_1, D_2) := \frac{|\textit{vocab}(D_1) \cap \textit{vocab}(D_2)|}{|\textit{vocab}(D_1) \cup \textit{vocab}(D_2)|}
    \]
    where $D_1$, $D_2$ are datasets and $\textit{vocab}(D_i)$ is the vocabulary-set of $D_i$, tokenized by whitespace.
    }
\end{itemize}

\PaperLong{
    We take a first step towards a similarity measure for entity matching benchmarks and use case datasets in Appendix~\ref{Appendix ssec: Analyzing the Correlation between Dataset Profiling Metrics and Matching Performance} by measuring the impact of several of the above factors on matching performance.
}

\subsection{Measuring Data Matching Quality}
\label{ssec: Measuring Data Matching Quality}

When ground truth annotations are available, a multitude of different metrics can be calculated.
While some are generally used and considered essential, others suit specific needs.
To be universally useful but highly adaptable, \frost focuses on many well-known metrics, but can be extended easily by any other metrics.
We distinguish between pair-based metrics and cluster-based metrics.

\subsubsection{Pair-based Metrics}
\label{sssec: Pair-based Metrics}

To compare an experiment $E$ against a ground truth annotation $G$ of a dataset $D$ as sets of pairs, the confusion matrix can be defined as shown in Figure~\ref{table: Confusion Matrix}.
\begin{figure}[t]
\centering 
\begin{tabular}{m{1.6cm} | m{2.6cm} | m{2.6cm} | } 
    & Positive & Negative \\ 
\hline
Predicted Positive & $E \cap G$ \hfill (TP) & $E \setminus G$ \hfill (FP)\\ 
\hline
Predicted Negative &$G \setminus E$ \hfill (FN) & $(\TwoElemSubsets{D} \setminus E) \setminus G$ \hfill (TN)\\ 
\hline
\end{tabular}
\captionWithDescription{Confusion Matrix}{Comparison of experiment $E$ against ground truth annotation $G$ on dataset $D$ as sets of pairs.}
\label{table: Confusion Matrix}
\end{figure}

This matrix allows the calculation of all metrics known from the context of binary classification.
Pair-based metrics do not require the identity link network of experiment $E$ to be transitively closed.
Therefore, they can be used to calculate matching quality even at intermediate stages of the matching pipeline.
For example, pair-based metrics allow measuring the performance of the candidate generation phase.
Additionally, they directly contrast the quality of matching solutions that return clusters with matching solutions that return pairs (and do not necessarily output transitively closed identity link networks)~\cite{Weis2006ADD}.
Note that pair-based metrics implicitly give more weight to larger clusters, as each pair of records within a cluster is counted towards the result.

Another weakness of pair-based metrics is the fact that in the real-world there is almost always a large imbalance between true positives and true negatives (called class imbalance)~\cite{Christen2012DataMatching}.
While a dataset of $n$ tuples usually contains only $O(n)$ duplicate pairs, it may consist of up to $O(n^2)$ non-duplicate pairs.
Metrics that judge upon correctly classified non-duplicates (true negatives) are therefore considered unreliable.
For example, the \emph{accuracy} of matching results compared to a ground truth might be close to~1, even when all record pairs were classified as non-duplicates.

\frost supports a wide selection of pair-based metrics considering the above observations including the common precision, recall and \fMeasure~\cite{Menestrina2010EvaluatingResults}, but also more special ones,
such as the Reduction Ratio~\cite{Koepcke2009FrameworksForEntityMatching}, the \fStarMeasure~\cite{hand2021FStar}, the Fowlkes-Mallows index~\cite{fowlkes1983folwkesMallowsIndex}, and the Matthews correlation coefficient~\cite{Chicco2021MCC}.

\subsubsection{Cluster-based Metrics}
\label{sssec: Cluster-based Metrics}

Cluster-based metrics are most often computed using similarities between clusters of the ground truth and the experiment~\cite{Menestrina2010EvaluatingResults,barnes2015practioners,Nanayakkara2019EvaluationMeasureRecordLinkage}.
An advantage of cluster-based metrics is that they are immune to the class imbalance described above.
On the other hand, they cannot be used to directly evaluate matching solutions that produce non-transitively closed sets of matches~\cite{Weis2006ADD}.
For example, the output of intermediate stages of a matching pipeline is usually not clustered. 

\frost utilizes several prominent cluster-based metrics including
the closest-cluster-\fMeasure~\cite{Benjelloun2009Swoosh}, 
the Variation of information~\cite{Meila2003VariationOfInformation}
and the Generalized merge distance~\cite{Menestrina2010EvaluatingResults}.

\subsubsection{Evaluating Quality Without Ground Truth}
\label{sssec: Evaluating Quality Without Ground Truth}
In many real-world use cases, labeled data is not available.
\frost also supports metrics and evaluation strategies that try to estimate matching quality on real-world datasets without ground truth annotations:

\begin{itemize}[leftmargin=.5cm]
\item{Idrissou et al.\ show that redundancy in identity link networks correlates with high matching quality~\cite{Idrissou2020NoGoldstandard}. Interestingly, their experiments show a ``very strong predictive power of […their] $e_\mathit{Q}$ metric for the quality of […identity link networks] when compared to human judgement''~\cite{Idrissou2020NoGoldstandard}.}

\item{The minimum number of pairs that must be added to or removed from the set of detected matches for it to be transitively closed is another relevant metric.
The larger this number, the more inconsistent the proposed matches.}

\item{Duplicate records are typically closer to each other than to other records.
Thus, the compactness of the individual clusters and the sparsity of their local neighborhoods as proposed by Chaudhuri et al.~\cite{GantiCM05} can estimate the quality of the whole matching result.
To calculate compactness and sparsity, however, we need similarity scores between the individual records provided by the matching solution for both matches (compactness) and close non-matches (sparsity).}

\item{If the set of detected matches is not transitively closed, we can achieve this closeness by applying several clustering algorithms~\cite{Hassanzadeh2009ClusterAlgorithms, Draisbach2019PairsToClusters},
such as maximum clique clustering or Markov clustering.
Here we can assume that the more similar the resulting clusterings are, the more consistent are the initially discovered matches.
Again, many clustering algorithms (e.g., Markov clustering) require similarity scores for the matches.}

\item{We can compare the matching result with those of other matching solutions applied to the same dataset.
The consensus on an individual matching decision (match or non-match) is a good indicator on its correctness~\cite{vogel2014ReachForGold}.
Thus, the total number of deviations from the majority votes can be used to estimate the quality of the whole matching result.}
\end{itemize}

Many of these aspects can be used not only to calculate a metric, but are predestined to guide users in the exploration of their matching results, e.g., by presenting record pairs that are likely misclassified by their solution (false positive or false negative). 
We describe these exploration techniques in Section~\ref{sec: Exploring Data Matching Results}.
\subsection{Soft KPIs: Effort and Cost}
\label{ssec: Measuring Data Matching Soft KPIs}

Every matching solution has different advantages and disadvantages and requires a different type of configuration.
As an example, supervised 
machine learning approaches need training data, whereas rule-based approaches need a set of rules.
When deciding which matching solution to use for a specific use case, these properties are of importance, as they influence how expensive and time-consuming it is to employ the solution.
To assist the decision process, \frost includes a benchmark dimension for soft key performance indicators (KPIs), which models such business aspects.

The main goal of these soft KPIs is to provide users a comparable overview of relevant, non-performance properties of matching solutions and experiments.
Most of these KPIs model the human effort (i.e., the amount and complexity of work) necessary to perform a specific task.
While many non-effort KPIs are objective and therefore easily comparable, effort is subjective and has to be estimated.
People with varying skills often have different opinions on how long it takes to configure a matching solution.
Therefore, we measure such effort using two variables:
\begin{inparaenum}[(i)]
\item The amount of time an expert needs to finish the task (HR-amount), and
\item the expert's skill level from 0 (untrained) to 100 (highly skilled).
\end{inparaenum}
HR-amount and expertise are interdependent.
When comparing two persons with different expertise, usually, the person with more expertise is faster.
Chatzoglou and Macaulay state that low experience is an indicator for increased time and cost, and that experience is considered an important factor for productivity~\cite{Chatzoglou1997Experience}.
Expertise is typically related to pay level.
Therefore, combining HR-amount and expertise yields a rough estimation of the monetary cost of performing the task.

The soft KPIs supported by \frost can be categorized into three classes:
\begin{itemize}[leftmargin=.5cm]
 \itemWithTitle{Lifecycle Expenditures}{
One important business aspect is the expenditure for integrating and operating a matching solution over its entire life-cycle.
Based on life-cycle cost analysis (LCCA)~\cite{2007EllisLifeCycleCost}, \frost supports several soft KPIs to represent the different product phases,
such as the general costs of the life-cycle
or the effort required to
get the matching solution ready for production within a company's ecosystem and configure the matching solution for its particular use case,
where we distinguish between domain-specific configurations (e.g., the manual labeling of training data)
and technique-specific configurations (e.g., the selection of algorithms). 
}

\itemWithTitle{Categorical Soft KPIs}{	
Apart from lifecycle expenditures, there are a few more aspects relevant for businesses:
These include the 
\begin{inparaenum}[(i)]
\item development types (e.g., on-premise or cloud-based),
\item interfaces (e.g., GUI, API, CLI), and
\item techniques (e.g., rule-based, clustering, or probabilistic decision models)
\end{inparaenum}
supported by the given matching solution.
}

\itemWithTitle{Soft KPIs of Experiments}{	
\frost supports measuring and evaluating soft KPIs on a per experiment basis.
This includes the effort needed to 
set up the experiment (e.g., acquisition of suitable test data) 
and the runtime that the matching solution required to complete the experiment.
}
\end{itemize}

The underlying effort and cost values need to be provided by the users.
However, \frost helps manage these numbers beyond single experiments
and supports calculating, comparing and evaluating all the aforementioned soft KPIs that are based on these numbers.
\frost supports two different evaluation techniques for soft KPIs. 
On the one hand, it provides a decision matrix including all above metrics side by side.
Importantly, this decision matrix also includes quality metrics to provide a holistic view of the attractiveness of the compared solutions.
On the other hand, \frost provides users the ability to aggregate metrics.
For example, to estimate costs, the effort-based metrics can be converted into costs as described above and added to general costs.
Because this aggregation depends on the use case, \frost does not pre-define aggregation strategies, but provides a framework for aggregating soft KPIs and quality metrics into use case specific KPIs.

As proposed and used by Köpcke et al.~\cite{Koepcke2009FEVER, Koepcke2010UsingFEVER}, \frost aids users in analyzing soft KPIs for experiments with a diagram-based approach.
This helps answer questions, such as how much effort is needed to achieve a specific metric threshold (e.g., 80\% precision), whether increased runtime yields better results, or how good a matching solution is out-of-the-box versus how much effort it takes to improve the results.
The diagram is especially interesting when experiments from multiple matching solutions are compared.
Evaluations thereby become competitive and allow discovering different characteristics of the matching solutions.

\section{Exploring Data Matching Results}
\label{sec: Exploring Data Matching Results}

The general workflow for improving matching solutions and arriving at a sufficient configuration is usually iterative.
Thus, after one run has finished, its results need to be analyzed to gain insights about the solution's behavior.
Afterwards, the matching solution can be refined accordingly and re-run.
As motivated in Section~\ref{ssec: Formal Matching Process}, we present structured approaches to explore data matching results.
Specifically, we reduce the amount of information presented to the user by filtering out irrelevant data, sorting it by interestingness, and enriching it with useful information about the type of error.
Finally, we introduce diagram-based evaluations.

\subsection{Set-based Comparisons}
\label{ssec: Set-based Comparisons}

Manual inspection of experimental results can be a poor experience.
As an example, some output formats consist solely of identifiers and thus require to be joined with the dataset to be helpful.
Additionally, only limited information can be extracted by looking at results side-by-side; in practice usually more than two result sets are compared.
A common use-case is to contrast multiple runs of the same matching solution with each other, or to evaluate differences between two distinct solutions and a ground truth. 

\frost supports a generic set-based approach to result evaluation that enriches identifiers with the actual dataset record. 
The set operations \emph{intersection} and \emph{difference} can describe all partitions of the confusion matrix, as introduced in Section~\ref{sssec: Pair-based Metrics}.
As an exemplary evaluation, consider two result sets $E_1, E_2 \subseteq \TwoElemSubsets{D}$, where $E_2$ serves as ground truth.
The subset of false positives is defined as the set of elements in $E_1$ that are not part of the ground truth $E_2$, or simply $E_1 \setminus E_2$. 
While the confusion matrix is limited to evaluating binary classification tasks with two result sets, the generic approach can compare multiple result sets.

As an intuitive visualization technique, \frost makes use of Venn diagrams.
When $n$ experiments are compared, these diagrams describe all $\binom{n}{2}$ possible subsets visually.
A disadvantage with Venn diagrams is that they get very complex for larger numbers of sets.
Venn diagrams of more than three sets need to use geometric shapes more advanced than circles~\cite{Ruskey2006SimpleSymmVenn}.
Set-based comparisons and Venn diagrams in particular can help to answer a variety of evaluation goals, such as:

\begin{itemize}[leftmargin=.5cm]
    \item{Compare two matching solutions' result sets against a ground truth to discover common pairs. This evaluation can easily be visualized with circle-based Venn diagrams.}
    
    \item{Find shortcomings or improvements of a new matching solution compared to a list of proven solutions by selecting all duplicate pairs only the new solution detected.}
   
    \item{Create an experimental ground truth~\cite{vogel2014ReachForGold} from the intersection of multiple experiments.}
\end{itemize}

Because exploration is supposed to be interactive, an implementation should provide vivid Venn diagrams. Clicking on regions should allow selecting the corresponding set intersection. Thereby, the desired configuration can be composed easily according to its visual representation.

\subsection{Pair Selection Strategies}
\label{ssec: Selection Strategies}

While set-based comparisons are useful on their own, real-world datasets can contain millions of records, making it unfeasible to examine all pairs in a set.
Therefore, strategies to reduce the number of pairs shown are crucial.
\frost supports a wide range of selection techniques to highlight relevant pairs which can be used separately or as a composition according to the current use case.

\subsubsection{Pairs around the Threshold}
\label{sssec: Pairs around the Threshold}

For matching solutions that provide a meaningful similarity threshold, an easy section of the result to further investigate is located close to the similarity threshold, as it includes information on border cases. Pairs in this section are usually considered uncertain, as a slight shift of the threshold may change their state. Nevertheless, they still yield helpful insights about what is especially difficult for the matching solution.
To select $k$ pairs, one can either choose $\frac{k}{2}$ pairs above and below the threshold or based on a certain proportion. For instance, one interesting proportion is the ratio of incorrectly classified pairs above the threshold to below the threshold.

\subsubsection{Incorrectly Labeled Outliers}
\label{sssec: Incorrectly labeled Outliers}

Another group of interesting pairs lies further away from the threshold.
For example, one could evaluate why the matching solution failed by searching for a common ``misleading'' feature among the selected pairs. 
Therefore, we allow selecting incorrectly labeled pairs that are the furthest away from the threshold.

\subsubsection{Percentiles with Representatives}
\label{sssec: Percentiles with Representatives}

Sometimes, the goal is to get an overview over the matching quality before diving into details. For this, we support finding representative pairs from all parts of the result set.
Conceptually, this strategy sorts result sets by a similarity score and then splits them into smaller partitions.
Each of these partitions is then reduced to a few representative pairs that represent the matching solution's behavior within this partition.

Let $E$ be a result (sub)set with $m$ pairs that is split into $k$ equally-sized partitions.
To sample $b$ representative pairs for each partition, different choices exist:
\begin{itemize}[leftmargin=.5cm]
    \itemWithTitle{Random sampling}{$b$ pairs are sampled randomly from each partition. While this technique is unbiased, it may also only yield uninteresting pairs and thereby no helpful insights.}
    
    \itemWithTitle{Class-based sampling}{For a partition with $k_\mathit{T}$ correctly and $k_\mathit{F}$ incorrectly classified pairs, we randomly sample
    $b \cdot k_\mathit{T} / (k_\mathit{T} + k_\mathit{F})$ correctly and
    $b \cdot k_\mathit{F} / (k_\mathit{T} + k_\mathit{F})$ incorrectly labeled pairs. Thereby, we make sure to weigh the numbers of pairs according to the algorithms performance.}
    
    \itemWithTitle{Quantile sampling}{Alternatively, $b$ pairs can be sampled by selecting $b$ quantiles, again based on the similarity score. For $b=5$, this would mean to select quantiles 0, 0.25, 0.5, 0.75, and 1. This technique has the advantage of unbiasedly representing the different parts of the partition.}
\end{itemize}

Additionally, we can label each partition with its confusion matrix and metrics.
Thus, users can focus on those partitions with high error levels.
A partition with few to no incorrectly labeled pairs is considered to be a confident section.
In contrast, a section with many false positives and/or false negatives is very unconfident, and therefore deserves more attention.

\subsubsection{Plain Result Pairs}
\label{sssec: Plain result pairs}

As outlined in Section~\ref{ssec: Formal Matching Process}, \frost requires result sets to be transitively closed.
On the one hand, this can lead to more realistic metrics.
But on the other hand, it can also enlarge small result sets to a very large number of pairs and thereby possibly introduces a substantial number of false positives.
Thus, \frost includes a selection strategy that will hide all pairs that were added by a clustering algorithm from a given result subset.
What remains are all pairs that were originally labeled by a matching solution.
To enable this, \frost requires information on which pairs were added during the clustering process and which were labelled by the matching solution itself.

\subsection{Sorting Strategies}
\label{ssec: Sorting Strategies}

Besides reducing the result sets to smaller subsets, \frost also supports to sort pairs by their \emph{interestingness} within a given subset. When relevant pairs are shown first, developers can gain insights more quickly to improve the matching solution's performance on a given dataset. The usefulness of the sorting procedure varies between strategies and use case. Below, we discuss several measures of interestingness of record pairs.

\subsubsection{Similarity Score}
\label{sssec: Similarity Score}

A common score to rank any set of pairs is the similarity of a pair's records. Whenever similarity values are available for all pairs, this technique offers a view on the data from the matching solution's perspective.

\subsubsection{Column Entropy}
\label{sssec: Column Entropy}

We also define independent scores that were not part of a matching solution's output.
For each token $t$ within a given cell, let $\textit{prob}_t$ be its occurrence probability within the cell and $\textit{columnProb}_t$ the probability within the column.
The cell entropy is calculated by:
\[
\sum\nolimits_{\textit{token}\:t}{\textit{prob}_t \cdot -log(\textit{columnProb}_t)}
\]
where the second factor describes a token's information content within its column. This formula is close to the original definition of entropy by Shannon~\cite{shannon1948mathematical}, but is applied column-wise.
For a given pair $p=\{r_\mathit{1}, r_\mathit{2}\}$, we can calculate its entropy as the sum of all cell entropies of both records. 
Pairs with a particularly high entropy score contain many rare tokens and are therefore expected to be easier to correctly classify. Depending on dataset and matching solution, we may observe a divergence in the distribution of entropy among the confusion matrix. If not, we can still use entropy as a score to sort pairs within a subset of the result set(s).

\subsection{Error Analysis}
\label{ssec: Error Analysis}

To better understand why a pair was misclassified by a certain matching solution, one could analyze why a similar pair was labelled correctly. Thereby, one can gain insights on why the matching solution came to a false conclusion and find errors within the decision model.
\frost allows enriching a misclassified pair~$p_\mathit{f}=\{e_\mathit{f,1}, e_\mathit{f,2}\}$ with a correctly classified pair~$p_\mathit{t}=\{e_\mathit{t,1}, e_\mathit{t,2}\}$.
We search for $p_\mathit{t}$ by considering only correctly classified pairs and selecting the one which is most similar to $p_\mathit{f}$.
We describe the similarity between the pairs $p_\mathit{f}$ and $p_\mathit{t}$ with vectors 
\[
\textbf{v}_{direct} = \colvec{sim(e_\mathit{f,1}, e_\mathit{t,1})}{sim(e_\mathit{f,2}, e_\mathit{t,2})}\text{ and }\textbf{v}_{cross} = \colvec{sim(e_\mathit{f,1}, e_\mathit{t,2})}{sim(e_\mathit{f,2}, e_\mathit{t,1})}
\] 
To compare these vectors with each other, we convert each one into a scalar distance measure. For this, the Minkowski metric with $q \in [1,2]$ is used against $\vec{0}$ as the reference point:
\[
    \textit{distance}(\textbf{v}) = D(\textbf{v},\vec{0}) = (|\textbf{v}_1 - 0|^q + |\textbf{v}_2 - 0|^q)^{\frac{1}{q}}
\]
For $q=1$, this equals the Manhattan distance and for $q=2$ the Euclidean distance. It depends on the user to choose $q \in [1,2]$ depending on the use-case.
Finally, we define the distance score of $p_\mathit{t}$ against $p_\mathit{f}$ as
\[
 \textit{score} = \max\{\textit{distance}(\textbf{v}_{direct}), \: distance(\textbf{v}_{cross})\}
\]
Whichever candidate pair~$p_\mathit{t}$ scores highest is then selected.

To receive best results, all possible pairs should include a similarity score. Since this would require the matching solution to compare $O(n^4)$ values for a dataset of size $n$, a possible extension to \frost could be to calculate a simple distance measure for a set of promising pairs internally.

\subsection{Diagram-based Exploration}
\label{ssec: Diagram-based Exploration}

All strategies so far are for set-based comparisons and either limit the number of pairs shown (Section~\ref{ssec: Selection Strategies}), sort them (Section~\ref{ssec: Sorting Strategies}) or add additional information (Section~\ref{ssec: Error Analysis}). Here, we introduce a set of diagrams that aid in understanding a matching solution's behavior.

\subsubsection{Metric/Metric Diagrams}
\label{sssec: Metric-Metric Diagrams}

\begin{figure}[t]
\centering
\includegraphics[width=\columnwidth]{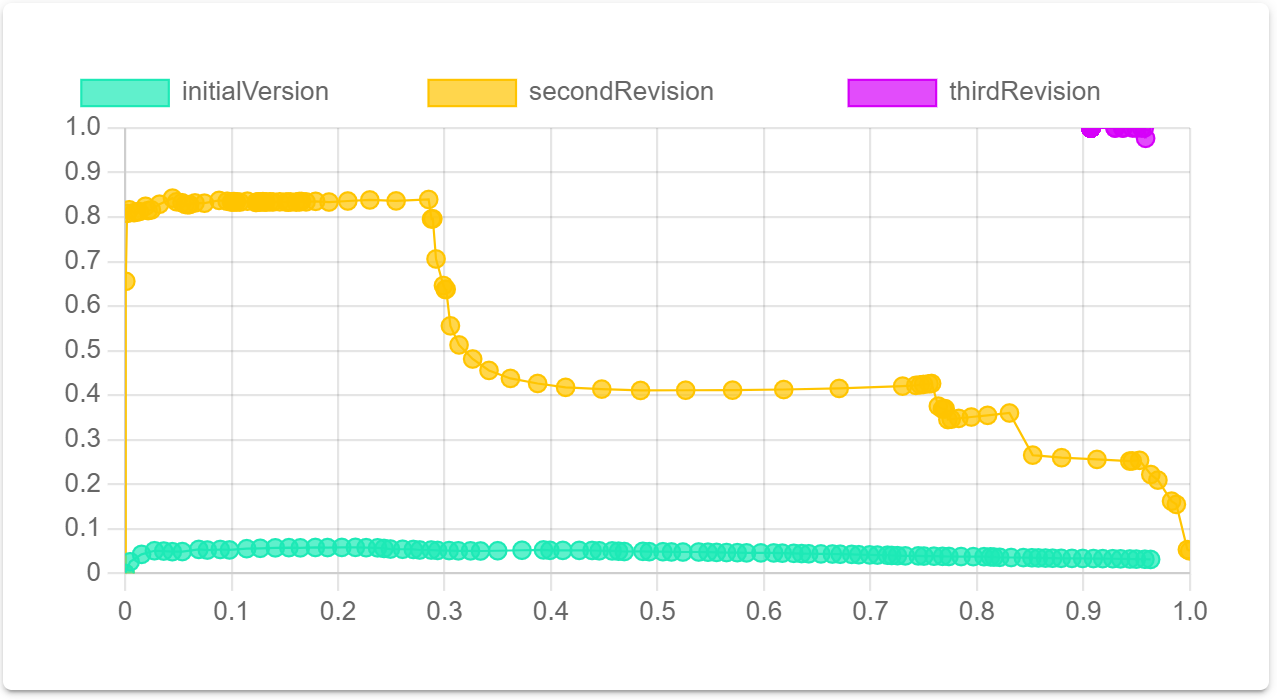}
\captionWithDescription{Precision-Recall Curve}{This diagram, taken from our implementation of Frost (Snowman, Section~\ref{sec: Reference Implementation}), plots recall against precision for a given set of similarity thresholds.}
\label{fig:snowmanSimilarity}
\end{figure}

For matching solutions that return similarity scores, one objective is to find a good similarity threshold.
\frost utilizes metric/metric diagrams for this objective.
Those diagrams compare two quality metrics against each other for a given set of similarity thresholds, and each data point is based on a different similarity threshold.
A commonly known diagram is the precision/recall curve (see Figure~\ref{fig:snowmanSimilarity}).
With it, one can visually observe which point (and thereby similarity score) yields the best ratio between both metrics.
Another well-known diagram is the ROC curve~\cite{gonccalves2014roc} plotting sensitivity (\emph{also:} recall) and specificity against each other.
This may not be suitable for every use case, though, as specificity depends on the count of true negatives.
However, depending on the shape of the curve, these diagrams may yield insights upon both a good similarity threshold and the reliability of the matching solution.
Next to using metric/metric diagrams in isolation, multiple diagrams between multiple matching solutions can be compared for competitive insights.

A limitation of this technique is that it heavily depends on how many pairs have a similarity score assigned.
In practice, metrics sampled at similarity scores significantly lower than the similarity threshold of the matching solution may not be representative because pairs with such low similarity scores are often excluded from the result set.

\subsubsection{Attribute Sparsity}
\label{sssec: Attribute Sparsity}

As missing attribute values are known to influence and complicate matching tasks \cite{crescenzi2021alaska,Petrovski2016SparseData,Primpeli2020ProfilingMatchingTasks}, we want to further investigate which attributes precisely affected a matching solution's performance most. 
Attribute Sparsity as introduced by Cres\-cenzi et al.\ measures how often attributes are, in fact, populated within a given dataset~\cite{crescenzi2021alaska}.
Thereby, the authors profile a given dataset's difficulty together with additional profiling dimensions. 
Since we aim to profile a matching solution's result set instead, we define a metric that measures the influence of null-valued attributes by the count of incorrectly assigned labels as follows:
Let $D$ be a dataset and $a$ be an attribute of $D$. 
We define $\textit{nullCount}(a)$ as the count of pairs in $\TwoElemSubsets{D}$ where at least one record of the pair is null in attribute $a$.
Additionally, we define $\textit{falseNullCount}(a)$ as the count of incorrectly classified pairs in $\textit{nullCount}(a)$ and $\textit{nullRatio}(a)$ as:
\[
    \textit{nullRatio}(a) = \frac{\textit{falseNullCount}(a)}{\textit{nullCount}(a)}
\]

In contrast to the raw $nullCount$, immoderate amounts of null occurrences within an attribute do not bias the $nullRatio$.
Calculating the metric for all attributes $a$ in $D$ yields a statistical distribution.
Graphical representations that show scores for discrete buckets, such as bar charts, support comparing measured scores. Thereby we can observe the following:
attributes with high $\textit{nullRatio}$ scores are statistically highly relevant for the matching decision as their absence could be related to many incorrectly assigned labels~\cite{Primpeli2020ProfilingMatchingTasks}.
For instance, we observed that the ratio reveals a high significance for the attributes \emph{author} and \emph{title} in the Cora dataset~\cite{Draisbach2010DuDeTD} for the Magellan matching solution~\cite{10.14778/2994509.2994535}.

If the revealed significant attributes do not match the expectation, this likely comes down to one of two reasons:

\begin{itemize}[leftmargin=.5cm]
    \itemWithTitle{Semantic mismatch}{
        A semantic mismatch exists if the matching solution weighs attributes  heavily that are semantically irrelevant for a matching decision. 
        For instance, a matching solution learned to weigh attributes $b$ and $c$ more significantly while $a$ and $b$ are more important in reality. 
        A semantic mismatch is an indication that the provided rule set or the learned network's weights are not consistent with the domain of the given dataset.
    }
    
    \itemWithTitle{Material mismatch}{
        A material mismatch exists if the statistically assumed significance of attributes is not adequate for the underlying dataset. 
        For instance, a matching solution weighs attributes $a$ and $b$ while the underlying dataset is often null in these attributes. 
        This mismatch might occur when a matching solution is used on another dataset than it was initially optimized for (for example due to transfer learning).
    }
\end{itemize}

A downside is that $\textit{nullRatio}$ relies on interspersed null values within the dataset $D$ and a meaningful and sophisticated schema.
Such a schema contains several attributes that provide meaningful information, for example street and city split instead of combined in a single address field.
For instance, the Cora dataset fulfills the requirements with an average attribute sparsity of $0.58$ and a schema with 17 attributes~\cite{Draisbach2010DuDeTD}. 

In conclusion, the exploration of $\textit{nullRatio}$ allows insights into the matching solution's handling of null values. 

\subsubsection{Attribute Equality}
\label{sssec: Attribute Equality}

Similar to attribute sparsity, \frost allows investigating the influence of equal attribute values on the matching process, too.
Equal attribute values can indicate a duplicate pair, although equality in one attribute is usually not a sufficient criterion.
For instance, while attributes, such as the person's name, may be sufficient for a match, others, such as post code, may not.
Therefore, \frost includes attribute equality as a dimension to statistically analyze which equal attributes are related to incorrectly assigned labels significantly often. 

Let $D$ be a dataset and $a$ be an attribute of $D$.
First, we define $\textit{equalCount}(a)$ as the count of pairs in $\TwoElemSubsets{D}$ where both records of the pair are equal in $a$. 
Second, we define $\textit{falseEqualCount}(a)$ as the count of incorrectly classified pairs in $\textit{equalCount}(a)$. We set:
\[
    \textit{equalRatio}(a) = \frac{\textit{falseEqualCount}(a)}{\textit{equalCount}(a)}
\]

A high $\textit{equalRatio}(a)$ for a given attribute $a$ indicates that the matching solution did not weigh the matching sufficiency of $a$ correctly (either too high or too low).
Again, calculating the metric for all attributes $a$ in $D$ yields a statistical distribution which, if compared across all attributes, can yield helpful insights. Similarly, bar charts can be used as an evaluation tool.

\section{Reference Implementation}
\label{sec: Reference Implementation}

\begin{figure}[t]
\centering
\includegraphics[width=\columnwidth]{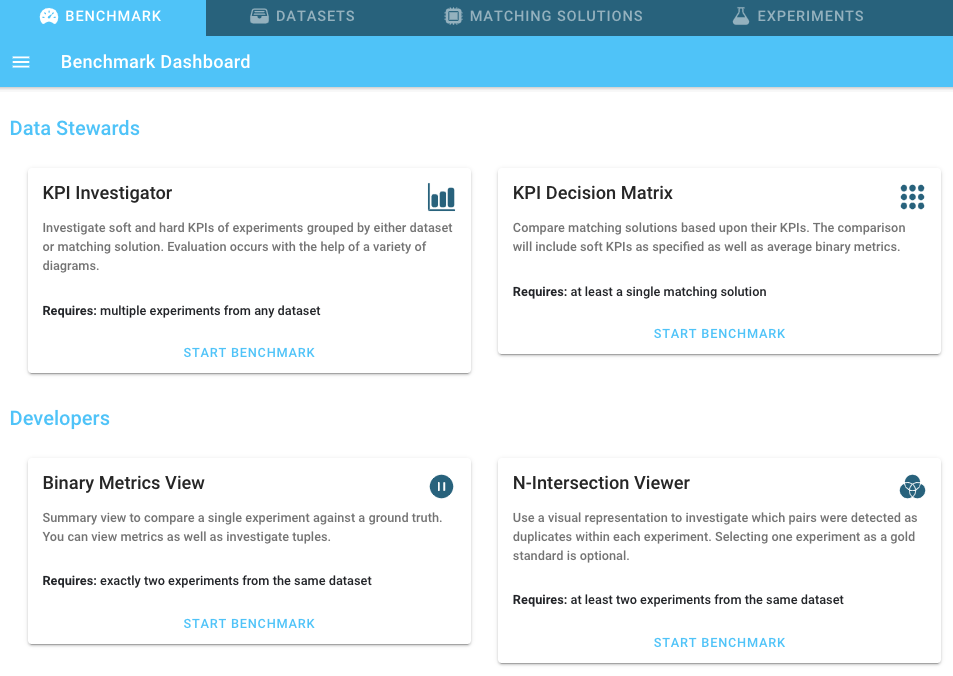}
\captionWithDescription{Snowman}{The start screen of the reference implementation of \frost: \snowman. From here, the individual benchmark actions are available.
}
\label{fig:snowmanApp}
\end{figure}

In this section, we present our reference implementation of \frost called \snowman and perform two example evaluations with it. Figure~\ref{fig:snowmanApp} shows a screenshot that highlights the fact that \snowman addresses both data stewards (domain experts) and developers.

\subsection{\Snowmans Features}
\label{ssec: Snowman Features}

\Snowman provides many features enabling developers to explore and evaluate data matching results. Next to evaluation techniques that require a ground truth, \snowman also supports those that operate solely on the matching results.

Besides traditional metric evaluation pages, \snowman has full support for our soft KPI dimensions from Section~\ref{ssec: Measuring Data Matching Soft KPIs} and supports the main exploration concepts from Section~\ref{sec: Exploring Data Matching Results}.
Below, we present a selection of evaluations that are already part of \snowman.
A full list can be found in \snowmans online documentation\footnote{\url{https://hpi-information-systems.github.io/snowman}}. 

\begin{itemize}[leftmargin=.5cm]
    \itemWithTitle{Data matching expenditures}{
        \snowman implements both the decision matrix and the diagram for evaluating experiment level expenditures as described in Section~\ref{ssec: Measuring Data Matching Soft KPIs}.
    } 
    
    \itemWithTitle{Set-based comparisons}{
        \snowman supports intersecting and subtracting experiments and ground truths with the help of an interactive Venn-diagram as described in Section~\ref{ssec: Set-based Comparisons} (see Figure~\ref{fig:snowmanIntersection}).
        To enhance the evaluation process, \snowman shows complete records instead of only entity IDs; if only intersection operators are used, clusters are grouped.
    }
    
    \itemWithTitle{Evaluating similarity scores}{
        \snowman helps users find the best similarity threshold by plotting the metric/metric diagrams discussed in Section~\ref{sssec: Metric-Metric Diagrams} (see Figure~\ref{fig:snowmanSimilarity}).
        It also allows to compare similarity functions of multiple matching solutions and multiple similarity functions of one matching solution.
    }
\end{itemize}

\Snowman provides a range of preinstalled benchmark datasets (including ground truth annotations) giving users the ability to easily evaluate and compare matching solutions in multiple domains without further imports.
Besides, it supports a range of different dataset and experiment formats and provides a convenient interface for additional custom CSV-based formats as well as other file-based experiment or dataset formats through customized importers.
Existing \href{https://hpi-information-systems.github.io/snowman/basic_usage/experiments/#import-formats}{importers for experiments} are 30-60 lines long and in the case of a CSV-based format as simple as defining the separator, quote, escape symbols and a mapping for rows to duplicate pairs or clusters.
\Snowman only requires the output that a matching solution provides. Further integration is not necessary.

\subsection{\Snowmans Architecture}
\label{ssec: Snowman Architecture}

\begin{figure}[t]
\centering
\includegraphics[width=0.8\columnwidth]{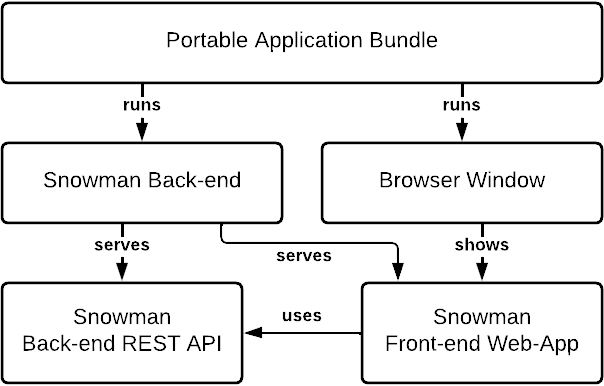}
\captionWithDescription{Architectural overview}{\Snowmans architecture is bundled into a single portable executable, but still offers a variety of extension points.
}
\label{fig:snowmanEasyArchitecture}
\end{figure}

Our application stack (see Figure~\ref{fig:snowmanEasyArchitecture}) makes use of \href{https://www.electronjs.org}{ElectronJS} and splits \snowman into a \href{https://nodejs.org}{NodeJS} \backend and a \href{https://reactjs.org}{ReactJS} \frontend.
Both are built using \href{https://www.typescriptlang.org}{TypeScript}, increasing maintainability and simplifying the onboarding process of new contributors.
More importantly, experiments show that many workloads are not significantly slower than for similar implementations in Java~\cite{BenchmarkJS}. 
\PaperShort{A more detailed architectural overview can be found in~\cite{frost2022arxiv}.}

All communication between \frontend and \backend occurs through a REST API specified according to the \href{https://www.openapis.org}{OpenAPI~3} standard.
This allows third-party applications to integrate easily with \snowman, for example to ingest matching results directly from within a matching solution or to automatically retrieve evaluation results, for example directly within Python3 code.
Furthermore, it means that \snowman can be deployed both locally and in a shared environment among multiple users.
To be easily usable in corporate environments where administrative privileges are rare and device settings might be restrictive, \snowman is portable and requires no installation or external dependencies.
All major operating systems are supported, including recent versions of Windows, macOS, Ubuntu, and Debian.

\subsection{\Snowmans User Experience}
\label{ssec: Snowman User Experience}

\begin{table*}[t]
\centering
\captionWithDescription{Runtime of Metric/Metric Diagrams}{The table shows a comparison of the runtime of \snowman's optimized algorithm for pair-based metric/metric diagrams against a na\"ive approach.
For each diagram, \numprint{100} different similarity thresholds were calculated.
}
\begin{tabular}{ l  r  r  r  r r}
\toprule
& & & \multicolumn{2}{c}{Metric diagram}  & \multirow[b]{1.9}{*}{\makecell{Approximate\\ speedup factor}}
\\ \cmidrule{4-5}
    Dataset
&
    Record count
&
    Matched pairs
&
     Custom
&
     Na\"ive
&
\\ \midrule 

    Altosight X4
&
    835
&
    \numprint{4005}
&
    184ms
&
    1.7s
&
    9
\\
    HPI Cora
&
    \numprint{1879}
&
    \numprint{5067}
&
    245ms
&
    7.4s
&
    30
\\
    FreeDB CDs
&
    \numprint{9763}
&
    147
&
    293ms
&
    16.4s
&
    56
\\
    Songs 100k
&
    \numprint{100000}
&
    \numprint{45801}
&
    1.6s
&
    43.9s
&
    28
\\
    Magellan Songs
&
    \numprint{1000000}
&
    \numprint{144349}
&
    6.1s
&
    6min 43s
&
    66
\\
\bottomrule
\multicolumn{6}{c}{}\\[-0.5em]
\end{tabular}
\label{Tab: Runtime of Back-end Algorithms}
\end{table*}

To be useful to enterprise data stewards, it is crucial for \snowman to provide results quickly for even the most demanding evaluations.
Most developers only have limited access to huge data centers, and gold standards containing multiple billion records are rare.
Therefore, we instead optimized \snowman so that it can run well on a typical enterprise laptop, but still provide results for medium-sized datasets in hundreds of milliseconds to a few seconds.\footnote{In case more compute power is required than available to the user on his local machine, \snowman can also be set up as a shared environment with its \backend hosted in the cloud.}

To achieve this, \snowman optimizes every step of the evaluation process, beginning with optimizing matching results while they are imported into the tool:
During import, a unique numerical ID is assigned to each record, allowing constant time access to records.
Additionally, a clustering of the experiment is constructed.
We do this because currently, nearly all calculations in \snowman are performed using transitively closed clusters instead of pairs, which leads to much faster runtimes (up to linear to the dataset length) in practice, compared to the quadratic number of pairs to be evaluated otherwise.
Let $D$ be the dataset that the matching solution was executed on, and let \textit{Matches} be a set of matches predicted by the matching solution.
The pre-calculations during the import of an experiment take $O(|\mathit{Matches}| \cdot log(|\mathit{D}|)$ \textit{(to map the dataset's native IDs to numeric persistent IDs)}.

Even when using clusterings instead of pairs to calculate results, in some cases, the run-time of evaluations on datasets with tens of thousands of records takes too long.
As an example, one of the most demanding evaluations of \snowman is calculating pair-based metric/metric diagrams (see Section~\ref{sssec: Metric-Metric Diagrams}).
A na\"ive approach to calculate these diagrams using clusterings is to sample metrics at different similarity thresholds without re-using insights from other thresholds.
Running this algorithm on a dataset with \numprint{100000} records and roughly \numprint{45000} matches produced by an industry-grade matching solution on an enterprise laptop led to a runtime of roughly 44~seconds (see Table~\ref{Tab: Runtime of Back-end Algorithms}), which is longer than most people comfortably wait for results.
To address this, we developed an efficient algorithm to compute metric/metric diagrams, which reuses intermediate results and dynamically builds a clustering while tracking relevant metrics.
An analysis of the algorithm shows that the worst-case runtime (excluding the time of the optimization during import) is in $O(|\mathit{D}| + |\mathit{Matches}| \cdot \mathit{s})$ where $\mathit{s}$ is the amount of data points on the diagram.
Additionally, the algorithm runs the faster, the more similar ground truth and experiment clusterings are.
Table~\ref{Tab: Runtime of Back-end Algorithms} confirms that, for real-world datasets, the algorithm is considerably faster compared to the na\"ive approach.
As an example, for the above-mentioned dataset with \numprint{100000} records and experiment with \numprint{45000} matches, it took only a little under two seconds.

In summary, \snowman enables users to run most evaluations directly on their laptops without the need for special hardware or a compute cluster while still enabling fast and easy iterations.

\subsection{SIGMOD Programming Contest}
\label{ssec: SIGMOD Contest}

The ACM SIGMOD programming contest 2021 presented the participants with an entity resolution task~\cite{Sigmod2021Contest}. 
The goal was to deduplicate three datasets and achieve the highest average \fMeasure.
All participants were given the opportunity to use \snowman as a pre-configured evaluation tool to investigate matching results.
After the contest finished, we analyzed five high performing matching solutions with our \benchmarkPlatform on the evaluation dataset $Z_4$. 
Three of the matching solutions used a machine learning approach, one used a rule-based approach, and one used a combination of rules and machine learning.
In the following, we present key insights uncovered by application of \snowman:

For an initial overview, we used \snowman's N-Metrics Viewer to compare quality metrics, such as precision, recall, and \fMeasure.
On average, the top-5 contest teams achieved an \fMeasure of $90.34\%$ with $87.4\%$ as the minimum and $92.7\%$ as the maximum. These results are impressive, as the dataset constitutes a quite difficult matching task: most of the matching has to be based on unstructured, cluttered information in the attribute \emph{name}.

As the performance of a matching solution is often strongly related to the selected similarity threshold, metric/metric diagrams as introduced in Section~\ref{sssec: Metric-Metric Diagrams} can be used to find the optimal threshold.
Using \snowman, we ascertained that two matching solutions had, in fact, not selected the optimal similarity threshold for their results.
Selecting a higher similarity threshold would have increased their \fMeasure by $8\%$ and $6\%$, respectively.
Surprisingly, these observations are also true for the training dataset.

With \snowman, we identified three true duplicate pairs that were not detected by at least four solutions.
This evaluation can be accomplished with the N-Intersection Viewer (see Figure~\ref{fig:snowmanIntersection}) by subtracting all result sets from the ground truth.
Interestingly, all three pairs include the record with ID \textit{altosight.com//1420}.
This is an indicator that this record is especially difficult to match, or all considered matching solutions make equal assumptions about related information representations.

These findings confirm that useful insights can be gained by coherently applying structured evaluation techniques and result exploration, emphasizing the need for a benchmark platform.
\PaperLong{In Appendix~\ref{Appendix ssec: Analyzing the Correlation between Dataset Profiling Metrics and Matching Performance}, we explore differences between the contest's test and training data.}
In summary, \snowman can help to better understand a matching task as well as result sets and thereby accelerates the development of successful matching solutions.

\subsection{Experimental Evaluation of Soft KPIs}
\label{ssec: Experimental Evaluation of Soft KPIs}

\begin{figure}[tb]
\centering
\includegraphics[width=\columnwidth]{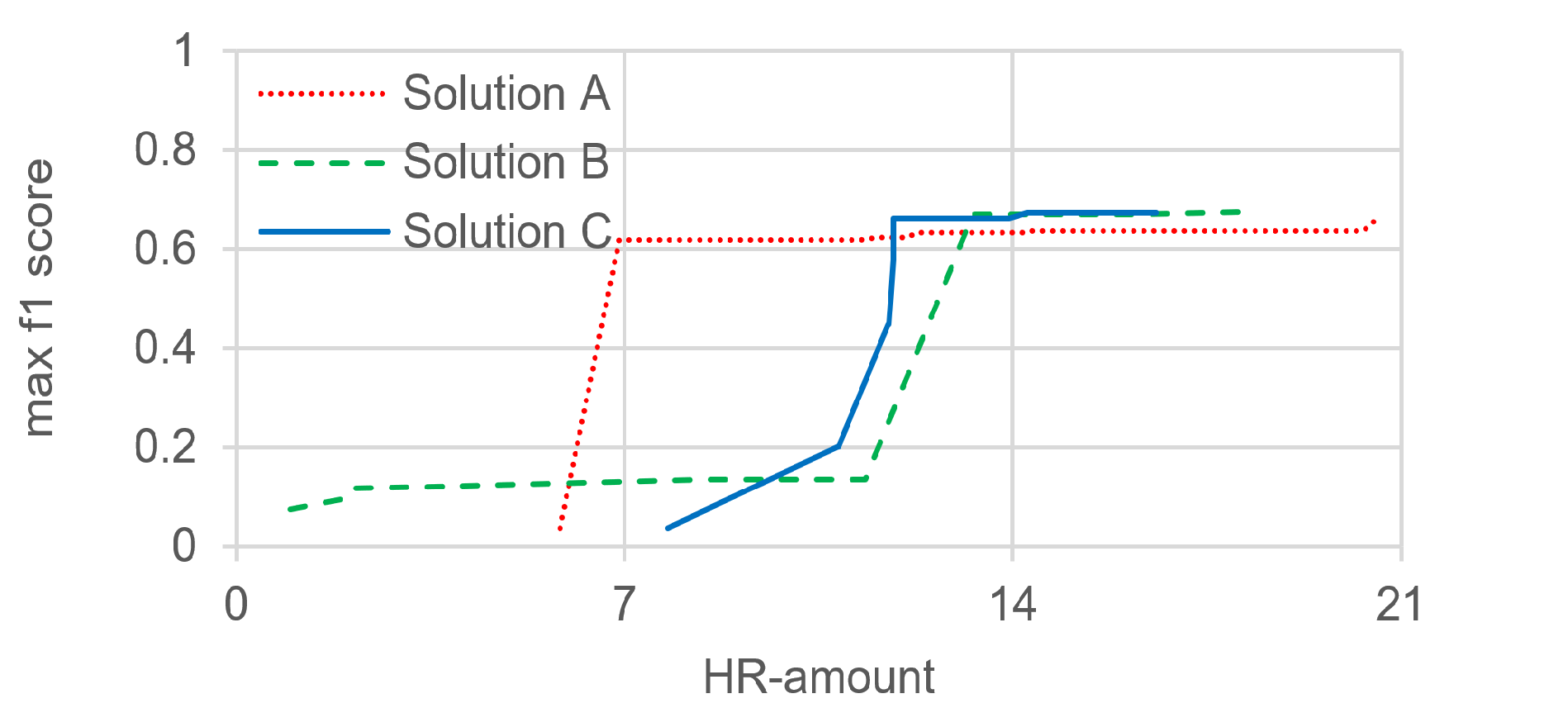}
\captionWithDescription{Maximum \fMeasure against effort spent (in hours)}{We optimized three solutions for the SIGMOD D4 dataset from scratch and tracked the effort spent throughout the process.}
\label{fig: Maximum f1 for different HR-amounts (in hours)}
\end{figure}

We conducted a study to show the relevance of effort measurements when benchmarking data matching solutions, and to examine what impact spent \emph{effort}, as discussed in Section~\ref{ssec: Measuring Data Matching Soft KPIs}, has on matching performance.
We expected that matching solutions improve with additional effort invested into their configuration, and that the curve of a target metric (e.g., \fMeasure) asymptotically approaches an optimum -- specific to each matching solution and dataset.
To validate this expectation, we manually optimized three different matching solutions, ranging from rule-based to machine learning approaches, for a given dataset.
Specifically, we deduplicated the SIGMOD contest's D4 dataset with the goal to optimize the \fMeasure achieved on the test dataset Z4 by using the training dataset X4 as well as its ground truth annotation.
Throughout the process, we tracked the effort spent.
Figure~\ref{fig: Maximum f1 for different HR-amounts (in hours)} illustrates how the \fMeasure evolved against the effort.

Each solution had a breakthrough point-in-time at which the performance increased significantly.
Afterwards, all solutions reached a barrier at around 14 hours, above which only minor improvements were achieved.
This could either mean that a major configuration change is required or that the maximum achievable performance for this matching solution on dataset D4 is reached.

\begin{figure}[t]
\centering
\includegraphics[width=\columnwidth]{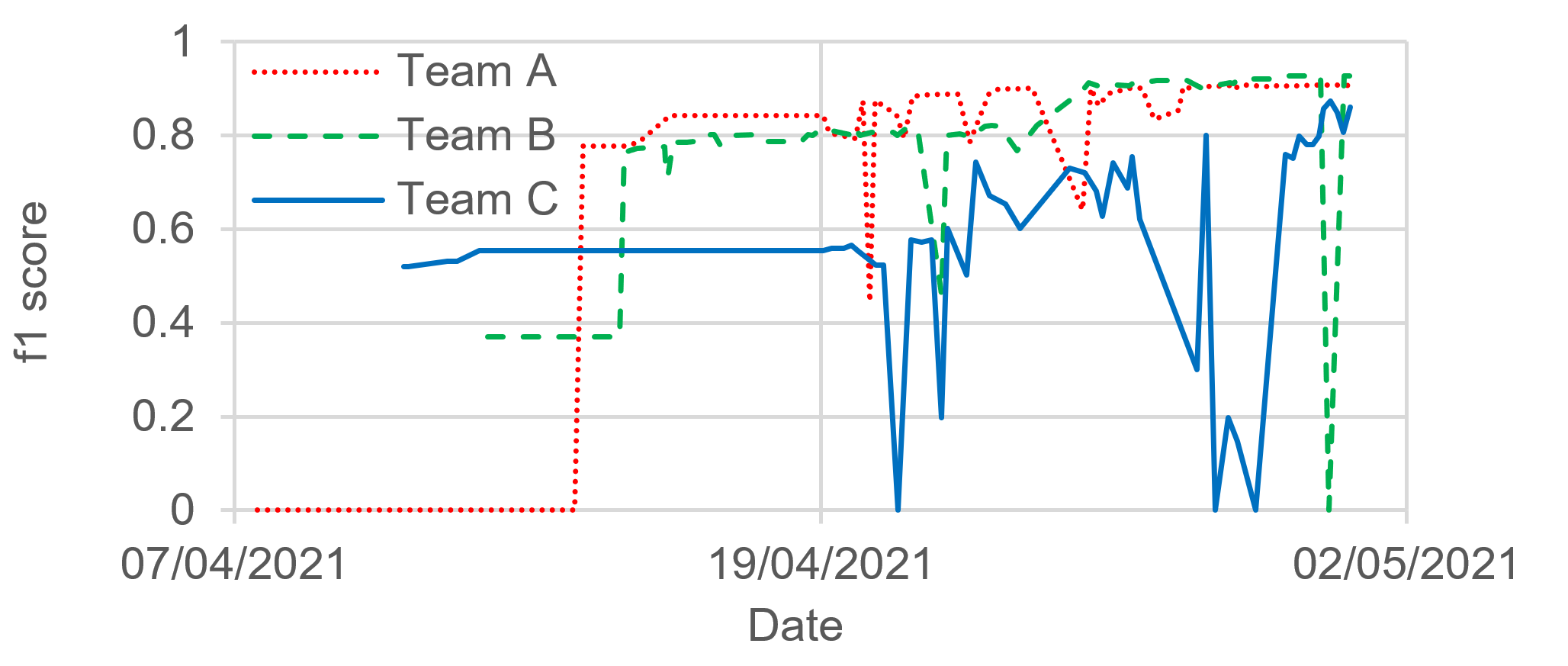}
\captionWithDescription{\fMeasure over time at the SIGMOD contest}{The evolution of the \fMeasure on dataset D4 of three of the top five teams at the SIGMOD contest.}
\label{fig: SIGMOD max f1}
\end{figure}

Additionally, we analyzed the \fMeasure of the submissions from five top teams of the SIGMOD contest over time (see Figure~\ref{fig: SIGMOD max f1}).
The matching quality of the different teams generally increased over time, but sometimes faced significant declines in matching performance.
Thus, the matching task had an overall trial-and-error character, which indicates that dataset~D4 seems to be challenging even for matching specialists. Furthermore, \frosts exploration features might reveal starting points for identifying the error of reasoning in taking new assumption for changing configuration. Thereby, \frost essentially foster business efficiency in configuring matching solutions.
In conclusion, effort diagrams are beneficial in a variety of ways:
They help users track the cost spent for optimizing matching solutions, to detect time points when larger configuration changes are necessary or additional effort might be wasted, prefigure when result exploration should be applied, and give insights about the difficulty of a dataset.

\section{Industry Use Cases}
\label{sec: Industry Use Cases}

Today's technologies offer a variety of approaches for data matching, targeted at a variety of use-cases.
Finding a well-performing matching solution is both crucial and difficult for companies facing poor data quality, resulting in non-transparent and suboptimal decisions.
Therefore, there is an industry-wide need for more transparency for data matching solutions.
In response, within SAP \frost has the goal to standardize the comparison and evaluation process for its various data matching solutions across departments.
Throughout this project, we received constant feedback from several teams within SAP working on future matching solutions.

In practice, matching solutions must be precisely adapted and optimized for a given use case.
For SAP customers and also for internal teams, this process is crucial to achieve maximum performance for their matching needs.
With the help of \frost's implementation \snowman, they were able to further optimize their workflow and receive helpful insights more quickly than with traditional, use-case specific evaluation tools.
Especially \snowmans ability to contrast multiple runs of a given matching solution against a ground truth allowed the teams to more quickly identify blind spots.

Beyond SAP, we were able to work with a global company on improving their in-house data matching solutions for business partner data.
With the help of \snowman, the company was able to evaluate alternative solutions to replace their current system in an interactive way. Thereby, information that would otherwise be only available to IT became directly accessible to data-matching stakeholders and domain experts:

\begin{itemize}[leftmargin=.5cm]
    \item \Snowman simplifies this process by providing a standardized way to compare matching solutions using hard- and soft KPIs.
    \item Additionally, revealing strengths and weaknesses became easier as \snowman conveniently displayed tuples found by both matching solutions, only one of them, or neither. 
    \item \Snowmans support for a variety of output formats makes this process easy and enables even persons with limited tech skills to gain meaningful insights.
\end{itemize}

\section{Conclusion and Outlook}
\label{sec: Conclusion and Outlook}

We introduced \frost, \aBenchmarkPlatform for data matching solutions.
Besides the traditional benchmark evaluation for result quality, it offers a dimension for expenditures as well as techniques to systematically explore and understand data matching results.
We examined how \frost can benefit both organizations in their buying decision and developers in improving their matching solutions.
Finally, we presented \snowman as a reference implementation for \frost and evaluated the results of this year's top teams from the SIGMOD programming contest with it.
Although we consider \frost and \snowman a significant step in the direction of a standardized and comprehensive \benchmarkPlatform for entity resolution systems, our long-term goal is to advance \frost even further along several lines of research:
\begin{itemize}[leftmargin=.5cm]
    \itemWithTitle{Compatibility with non-relational data}{Data matching is relevant beyond tabular data.
    Thus, \frost needs support for non-relational data models, such as XML or JSON.    
    } 
    \itemWithTitle{Selecting benchmark datasets}{
        As discussed in Section~\ref{sssec: Finding a Representative Benchmark Dataset}, it is difficult to find representative benchmark datasets for a real-world matching task. A suitability score based on profiling metrics would be an important contribution towards the search for suitable benchmark datasets.
    }
    \itemWithTitle{Categorizing errors}{
        The ability to categorize the errors of a matching solution helps to more easily find structural deficiencies.
        For example, a matching solution could be especially weak in the handling of typos.
    }
    \itemWithTitle{Recommending matching solutions}{
        A long-term goal might be to gather matching solutions, benchmark datasets, and evaluation results in a central repository.
        To assist organizations with real-world matching tasks, \frost could use this information to automatically determine promising matching solutions.
    }
\end{itemize}
\balance

\bibliographystyle{ACM-Reference-Format}
\bibliography{references}

\PaperLong{
    \clearpage
    \appendix
    \section{\Snowman's Run-Everywhere High-Performance \Backend}
\label{Appendix ssec: Creating a Run-Everywhere High-Performance Back-end for a Data Matching Benchmark in TypeScript}

Building the reference implementation \snowman\footnote{\url{https://github.com/HPI-Information-Systems/snowman}} for \frost is a challenging task, because the \benchmarkPlatform has to meet several conflicting requirements. Our final application stack makes use of ElectronJS and splits \snowman into a NodeJS \backend and a web\-app as its \frontend. In this section, we outline requirements and decisions that led to \snowman's application stack, as well as discuss its benefits and shortcomings.

The following list summarizes the key requirements that influenced \snowman's development:

\begin{enumerate}
    \item The application should be able to run on all major operating systems, including Windows~10, macOS~11 Big Sur as well as current versions of Ubuntu and Debian.
    \item Deployment should be possible both on local computers and in shared cloud environments.
    \item For local deployment, no installation or external dependencies should be required. Also, \snowman should be able to run without administrative privileges, as these are usually unavailable in commercial settings.
    \item Installation, upgrade, and removal procedures should be as simple as those of apps on a smartphone.
    \item \snowman should be easy to integrate with 3rd party applications via a http-based application programming interface (API).
    \item All functionality included within the \frontend should also be made available through the API.
    \item The tool should be interactive and answer usual requests in less than one second.
\end{enumerate}

\subsection{Architecture}
\label{Appendix sssec: Back-end Architecture}

\snowman's architecture is split into a \frontend and a \backend, which run independently and communicate via an API. Section \ref{Appendix sssec: Back-end REST-ful API} outlines \snowman's API interface in detail. As the \frontend uses an API to communicate with the \backend, it does not contain functionality that cannot be also accessed via \snowman's API\@. See Figure~\ref{fig:snowmanEasyArchitecture} on page~\pageref{fig:snowmanEasyArchitecture} for an overview of \snowman's architecture.

Both \frontend and \backend run on top of the ElectronJS\footnote{\url{https://www.electronjs.org}} portable application bundle, which allows the application to be shipped as a single binary file. As \snowman's \frontend is a web application, it is based on web technologies, such as HTML, CSS and JavaScript. The latter is also used in the \backend. To detect and prevent issues caused by loose typing, the TypeScript preprocessor\footnote{\url{https://www.typescriptlang.org}} is used to enforce strict type checking. See Section~\ref{Appendix sssec: Back-end Platform} for more details.

As \snowman describes a benchmark for relational data only, a relational database was chosen to persist data within the \backend. Relational databases allow for quick access to all records in different sorting orders using index structures. Normal database management systems (DBMS) require a separate process to install and operate.
To mitigate the need for external dependencies, we chose SQLite\footnote{\url{https://www.sqlite.org/index.html}} as our DBMS, which can be bundled together with the application.

\subsection{Platform}
\label{Appendix sssec: Back-end Platform}

ElectronJS is a toolkit to bundle web applications into standalone and multi-platform desktop apps. Each application consists of a \frontend webpage that runs inside a slimmed-down Chromium browser window, as well as a \backend process running on NodeJS (see Figure~\ref{fig:snowmanEasyArchitecture}). 
The chromium engine thereby isolates the web application from the host computer similar to a normal web browser. Privileged operations including access to system resources have to pass through the \backend process with inter-process communication. Accordingly, it remains to the \backend to sanitize and authorize requests received from the \frontend before passing them to the operating system.
After the application startup phase, \snowman does not make use of inter-process communication but rather passes all communication through the REST API.

The installation process is simple, as all program code resides within the downloaded artifact. To upgrade, one has to download a newer artifact and replace the old one. Similarly, uninstalling is handled by deleting the artifact and the application data.

By using ElectronJS, we benefit from the large ecosystem around JavaScript code.
In fact, JavaScript's package manager NPM is considered the largest ecosystem for open-source libraries in the world\footnote{\url{https://mirzaleka.medium.com/exploring-javascript-ecosystem-popular-tools-frameworks-libraries-7901703ec88f}}.
All functionality defined within the NodeJS standard library is platform-independent. This alleviates the need to write platform-specific code and amounts to cleaner code.

\subsection{Data Storage}
\label{Appendix sssec: Back-end Data Storage}

Although SQLite3 has many advantages as outlined above, it also comes with a major drawback.
Each database file needs to be locked before a write operation can occur.
During a lock-phase, other requests cannot read or write the target database file. Since \snowman's \backend is single-threaded, we do not consider this as a practical limitation.
Additionally, one could easily prevent this behavior by splitting the data into multiple database files through slight code changes.
To access the data efficiently, we developed an Object Relational Mapping (ORM). It constructs objects out of data retrieved from the database while enforcing strict typing. Compared to other ORMs, it was created precisely for our use case and is therefore faster and more flexible. Performance improvements are mainly achieved by a specific caching algorithm for our SQL statements. Additionally, this ORM enables the \backend to dynamically create schemata for each dataset or experiment on an existing connection.

\subsection{REST-ful API}
\label{Appendix sssec: Back-end REST-ful API}

As \snowman's \backend and \frontend communicate through an API, it has to be well-structured and properly documented.
We designed all routes according to the concept of Representational State Transfer~\cite{DBLP:conf/icse/FieldingT00} and documented them in an OpenAPI~3.0 compliant API specification\footnote{\url{https://www.openapis.org}}.
Besides an easy-to-understand API, we also gain the ability to automatically generate a compatible client and server based on the API specification.
Thus, correct and consistent types for our domain objects are maintained automatically across \frontend and \backend.
Generators\footnote{\url{https://github.com/OpenAPITools/openapi-generator}} exist for a variety of programming languages, which allows \snowman's users to easily integrate \snowman API into their own code.
For example, one could automatically upload results into a (potentially shared) \snowman instance after the code finished execution.
The latest API specification can be explored interactively as part of \snowman's documentation\footnote{\url{https://hpi-information-systems.github.io/snowman/openapi}}.

\subsection{Client-Server Version Mismatch}
\label{Appendix sssec: Preventing Client-server Version Mismatches}

All client-server applications face issues when the versions of client and server mismatch.
A traditional solution to this problem is to enforce versioning and have both client and server be backwards-compatible.

Instead, \snowman's client retrieves the \frontend web\-app from the \backend it is connected to -- no matter whether it is running locally as part of the bundle or remote on a server. Thereby, the \frontend presented to the user is always the same version as the \backend it is communicating with.
This has the additional advantage that users can also access the remote \frontend with a normal web browser.

\snowman's CLI faces a similar issue as it also makes use of \snowman API. As the Restish\footnote{\url{https://rest.sh}} CLI used is not specific to \snowman but rather a generic tool, it cannot face a version mismatch. Instead, it will retrieve the API specification before every request and adjust its behavior accordingly.

\subsection{Performance}
\label{Appendix sssec: Back-end Performance}

Besides interactive visualizations, \snowman's main goal is to analyze matching results quickly. Therefore, the \backend must allow for swift calculation of all necessary computations. 
Although JavaScript is not known for outstanding performance, experiments show that most workloads are not significantly slower than similar implementations in Java\footnote{\url{https://benchmarksgame-team.pages.debian.net/benchmarksgame/fastest/javascript.html}}.
In its current implementation, \snowman does not implement any parallel processing. Instead, NodeJS' default single-threaded event loop is used to handle all API requests, as we would otherwise have to synchronize manually. Still, most requests can be executed in a matter of milliseconds and long-running requests usually occur only for less usual imports and exports which are not part of the actual data analysis.

As a first step to profile \snowman's performance, we measured how long rare requests take. A typical import operation for datasets with fewer than \numprint{10000} tuples takes less than 1 second. The large Magellan Songs dataset with \numprint{1000000} records~\cite{dasXXXXMagellanData} requires about 1:15 minutes to complete its import. 

As \snowman is an interactive benchmark platform, evaluation operations need to be especially fast.
Thus, we also profiled \snowman's performance in the generation of metric/metric diagrams as introduced in Section~\ref{sssec: Metric-Metric Diagrams}.
The exact algorithm is outlined in Appendix~\ref{Appendix ssec: Constructing Dynamic Intersections for High Performance Precision/Recall Diagrams}.

We tested the performance with five different benchmark setups based on the Altosight X4 dataset from the SIGMOD contest 2021, the HPI Cora datset\footnote{\url{https://hpi.de/naumann/projects/repeatability/datasets/cora-dataset.html}}, the HPI FreeDB CDs dataset\footnote{\url{https://hpi.de/naumann/projects/repeatability/datasets/cd-datasets.html}}, a subset of \numprint{100000} songs from the Magellan Songs dataset as well as the whole Magellan Songs dataset. Table~\ref{Tab: Runtime of Back-end Algorithms} on page~\pageref{Tab: Runtime of Back-end Algorithms} shows our results on the computation time required for a Metric/Metric diagram. The measurements reveal a significant speed-up factor for our custom algorithm compared to the na\"ive approach (see Appendix~\ref{Appendix ssec: Constructing Dynamic Intersections for High Performance Precision/Recall Diagrams}). Even the largest dataset we tested requires only about $6.1s$ to finish the computation. Subsequent evaluations make use of caching, which yields an additional performance improvement. 

All tests were conducted with \snowman version 3.2.0\footnote{\url{https://github.com/HPI-Information-Systems/snowman/releases/tag/v3.2.0}} on a typical enterprise Windows 10 laptop with an Intel i5 quad-core processor 8th gen (Hyper-Threading enabled), 16 GB of DDR4 RAM and SSD storage. 
For fairness, both approaches were allowed to use index structures created during the initial import. 

Although very large datasets require significantly more processing time than mediums-sized ones, all major algorithms used to calculate evaluation results feature less than quadratic worst-case run time in the number of records.
In case the performance would become a bottleneck in the future, multiple options to cross-compile JavaScript code to native code exist. Also, computation could be split into multiple parts to enable multithreaded processing or outsourced to a dedicated \backend such as Apache Spark.

\subsection{Conclusion}
\label{Appendix sssec: Back-end Conclusion}

In conclusion, \snowman's architectural decisions were closely dictated by the requirements. Although other frameworks exist that could offer some of these features, the environment we chose is especially flexible and reflects the state-of-the-art. With the initial open-source release in March 2021, we took the first step towards building a community on GitHub. With public documentation\footnote{\url{https://hpi-information-systems.github.io/snowman}} and transparent development discussions, we hope to attract researchers and businesses world-wide to collaborate on \snowman so that it can improve and grow further.

\section{\Snowman's \Frontend Platform Architecture}
\label{Appendix ssec: Snowman's Front-end Platform Architecture}

\snowman addresses three groups of users: developers, researchers and data stewards. In particular, for data stewards, we do not assume experience in using a command line interface as provided by the \snowman back-end.
Thus, \snowman features a graphical user interface.
To simplify the onboarding of new users, one would ideally use fixed evaluation workflows that guide them throughout the evaluation.
The apparent solution is to have an independent linear workflow for each user group and thereby assume disjunct sets of evaluations for each of them.
In practice, these sets are not disjoint, but rather heavily overlap.
Therefore, evaluation tools need to be independent of particular personas.

Evaluations are configured by selected benchmark components. Benchmark components determine what is to be compared; matching solutions, experiments or datasets. 
Furthermore, evaluation opportunities are manifold as described in Sections~\ref{ssec: Exploration Opportunities (related)},~\ref{ssec: Measuring Data Matching Quality},~\ref{ssec: Measuring Data Matching Soft KPIs}, and~\ref{sec: Exploring Data Matching Results}. Throughout \snowman's lifecycle, additional evaluation strategies may become relevant.

In the end, \snowman requires to be a platform for different evaluation tools that receive their configuration from a generic configurator and operate independently of each other. 
The platform architecture acts as a runtime for evaluation and benchmark tools called sub-apps (e.g., Metric/Metric Diagrams from Section~\ref{sssec: Metric-Metric Diagrams}), encapsulates them into separate strategies, and provides them with common features.

\subsection{\Frontend Technology Introduction}
\label{Appendix sssec: Snowman's Front-end Tech Stack}

\snowman is based on ReactJS as it is a declarative and component-based library and thereby matches with the intuition of building a platform.
Due to the popularity of ReactJS among web developers\footnote{\url{https://insights.stackoverflow.com/survey/2020\#most-loved-dreaded-and-wanted}}, it is well suited for an open-source project that depends on future community contributions. 
The underlying architectural design pattern is mainly determined by the state management solution chosen. Besides the native ReactJS state features, there exist alternative libraries that are more adequate for complex applications\footnote{\url{https://kentcdodds.com/blog/prop-drilling}}, such as Redux\footnote{\url{https://redux.js.org}} and MobX\footnote{\url{https://mobx.js.org/README.html}}. 
\snowman uses Redux as it has great popularity and is used in many complex applications\footnote{\url{https://www.npmjs.com/browse/depended/react-redux}}.
Redux is based on the \emph{Flux} paradigm introduced by Facebook, but further expands the paradigm to make it more powerful. For instance, it describes a complex dispatcher structure which is constituted by reducers, as explained below. This additional complexity aids in the construction of complex applications such as \snowman.

\begin{figure*}[t]
\centering
\includegraphics[width=16cm]{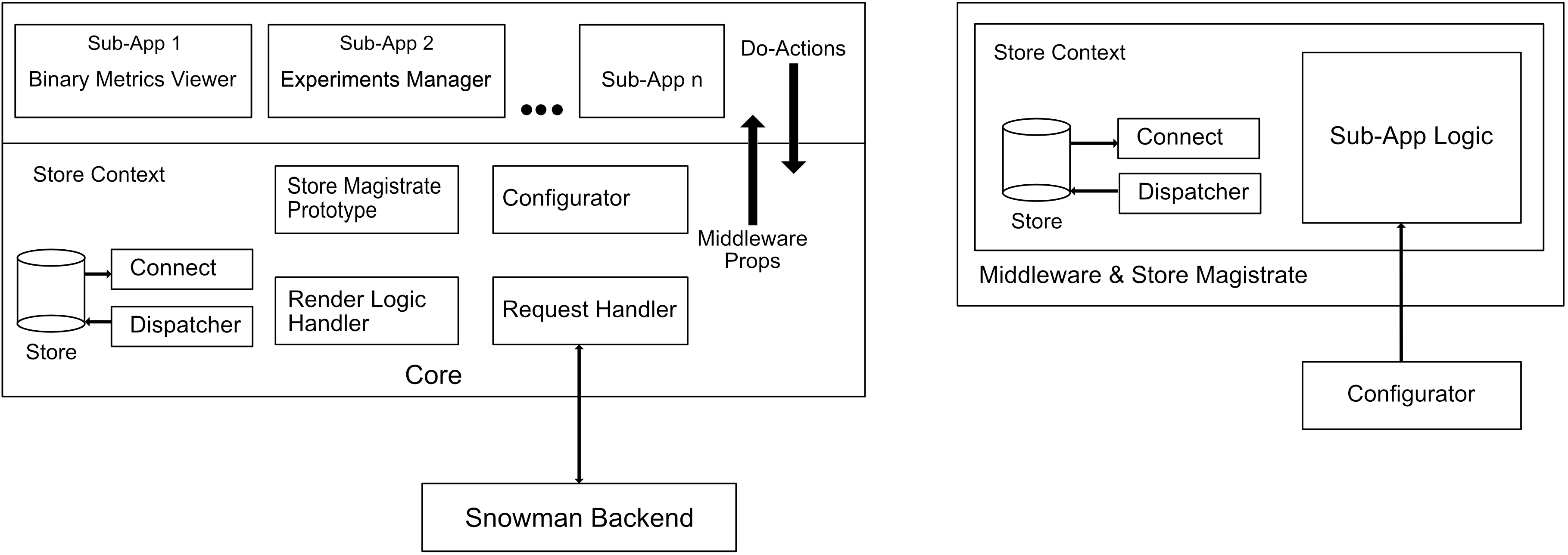}
\captionWithDescription{\snowman's \Frontend Architecture (left) and the structure of a sub-app (right)}{}
\label{fig:snowmanFrontendArchitecture}
\end{figure*}

Generally, a \frontend designed according to the Flux paradigm consists of view components that always reflect the state of the data. If the data changes, the view reacts to the state change.

\subsection{The Platform Architecture}
\label{Appendix sssec: The Platform Architecture}

Comparing different experiments on a single dataset with the help of quality metrics as introduced in Section~\ref{ssec: Measuring Data Matching Quality}, constitutes a significant workflow.
As this workflow represents a linear user story, chained reducers model it adequately: One chooses a target dataset as well as related experiments, and is then presented with applicable evaluation tools. Matching solutions thereby only serve as a filter criterion.

Additional benchmark views dock easily on to this linear workflow, but are limited to evaluate experiments of a single dataset. Thus, more complex resource selections, such as experiments from multiple datasets, are prohibited by this concept. 

Therefore, adding further user stories requires the implementation of multiple, co-existing linear workflows. 
However, missing data exchange between workflows would require users to configure their benchmark task for each workflow separately and thus multiple times. Additionally, as steps of different user stories cannot be combined in an obvious way, synchronized state may confuse users when switching between workflows.

As pure Redux demands a single store that contains the entire application state, further architectural issues arise: Each React component receives and accesses the entire application state. Thus, each component has to know an exhaustive set of details about the store. Beyond that, it also has to register new concerns at the root reducer.

To put it in a nutshell, a pure Redux architecture featuring linear workflows does not meet the platform requirements outlined above. 

As an alternative to the linear workflow, we introduce a new conceptual schema that allows a platform architecture.
As outlined in Figure~\ref{fig:snowmanFrontendArchitecture}, the \frontend consists of a common core as well as individual sub-apps. The core provides low-level logic to all sub-apps. For instance, it handles which sub-app is currently active and provides the http request handler. Each sub-app represents a benchmark or evaluation tool as well as tools to manage datasets, experiments and matching solutions.

A generic benchmark configurator is part of the core, too. It is able to model all possible configuration workflows: Based on the currently selected evaluation tool, adequate benchmark components can be selected. Limitations are defined by the evaluation tool itself. In contrast to the linear workflow, all required interpretations (i.e., which experiment is the ground truth) are thereby encoded within the configurator.

As part of our Redux architecture, we opted for each sub-app, as well as the core itself, to use separate Redux stores and only allow communication over a well-defined interface.
Since we thereby limit the amount of knowledge required by each sub-app to a minimum, we enable a true platform architecture within \snowman's \frontend.

\subsection{Technical Implementation}
\label{Appendix sssec: Technical Implementation}

Technically, a Redux store and its associated functionality work by extending ReactJS context methods. Context methods link values and methods defined outside into  the local component scope. Since it is possible to nest contexts, a component can only access resources from the context it is contained in, but not any overlying context.
Thus, nesting sub-apps by context encapsulates them and only gives them access to their own store defined within the current context.

In \snowman, a self-designed middleware abstracts the encapsulation process and declares a common interface between the core and the sub-apps.
Its responsibilities also include to carry out instructions from the core, for example to show the sub-app when it is active. 
Moreover, following the dependency injection pattern, control over store references has to be further abstracted and separated from the middleware. Additionally, as initialized Redux stores exist over the application's runtime and are not affected by garbage collection, \snowman requires a mechanism to keep track of its stores.

To address this issue, so-called store magistrates control pools of store instances. Once the sub-app is about to appear, the middleware generates an identifier and requests a store instance from the corresponding magistrate with it.
If the supplied identifier already exists within the pool, the magistrate reuses the existing store instance.
Otherwise, the store magistrate constructs a new store instance, pushes it to the pool, and returns its reference.

As a further optimization, we propose virtual stores, which represent slices of a materialized store and do not directly depend on a context.
Thus, magistrates can control them across their whole life-cycle, including destruction.
We created a reference implementation of virtual stores\footnote{\url{https://github.com/HPI-Information-Systems/snowman/pull/185}} that shows their greater flexibility.

\subsection{Discussion}
\label{Appendix sssec: Discussing the Platform-Grade Architecture}

The platform architecture introduced above has both advantages and disadvantages. In the following, we argue in favor and against \snowman's platform architecture and its implementation.

As outlined, the platform architecture comes with a two-sided benchmark configurator.
On the one hand, it encapsulates the concerns for selecting benchmark parts.
Thus, sub-apps do not have to take care of these concerns themselves.
Furthermore, additional configurators are not required because a single generic one can fit all possible workflows.
On the other hand, the configurator incorporates an inherent complexity for developers and is therefore more complicated to maintain and extend.
Additionally, we observed that end users might also be overwhelmed by its complexity and the missing guidance. Users require less support in understanding a linear, specific workflow compared to the general, standardized workflow. A UX expert confirmed these observations.
To counteract these effects, a guided introduction to the most important use cases of \snowman should be added.

Next, the de-structuring into sub-apps does not only yield advantages, but might also be problematic as most sub-apps still require \snowman as their runtime environment.
Thus, they cannot be used independently or would require substantial modifications to become independent.

Additionally, due to the inherent complexity and cohesion of the core, extracting a subset of its components and functionalities is difficult.
Furthermore, extensions and modifications to the core require in-depth knowledge about its structure as well as the component's choreographies.
For \snowman we do not consider the outlined disadvantages significant enough. Instead, advantages such as the core's small, easy to use core interface, and the fact that it is designed analogously to well-known and proven design concepts of operating systems, weigh more significant.

In conclusion, we therefore consider \snowman's platform-grade architecture to be a success. Nevertheless, it has to tackle the same issues as every platform-grade architecture:
Changes to the interface between core and sub-apps are critical, because every sub-app has to implement the new core interface upon the change.
Even though changes to the core itself are difficult, the abstraction becomes a game changer when new benchmark and evaluation apps are designed as they are truly independent of existing sub-apps.
We hope that \snowman gains additional popularity through the resulting extensibility and customizability.

\section{Correlating Dataset Properties and Data Matching Performance}
\label{Appendix ssec: Analyzing the Correlation between Dataset Profiling Metrics and Matching Performance}

The quality of a matching solution strongly depends on the dataset on which it is executed.
It might perform differently on a sparse dataset than on a dense dataset, for example.
A dataset with many non-atomic attributes presents a matching solution different challenges than a dataset with mainly atomic attributes.
Therefore, knowing the characteristics of the underlying dataset is important to develop high-performing matching solutions.

\frost can be used to investigate differences in matching behavior when running a matching solution on datasets with different properties.
An essential problem is that real-world scenarios usually do not feature ground truth annotations, which are necessary to evaluate matching solutions.
However, over the past few decades, 
the data matching community has compiled a set of annotated datasets that are typically used by researchers to develop and evaluate their matching solutions. Those datasets are called ``benchmark datasets''.
As described in Section~\ref{sssec: Reference Datasets}, some of these datasets are from real-world scenarios and have been annotated through laborious processes.
In contrast to researchers, practitioners usually do not want to evaluate the behavior of a newly developed matching solution, but must use these solutions to detect all duplicates within their use case-specific datasets.
Since these datasets do not provide a ground truth, a typical challenge for them is to find a benchmark dataset that is similar to their given use case dataset and thus can be used for evaluation instead.
This in turn requires the development of similarity metrics suitable to compare these datasets appropriately.

In the following, we analyze how different dataset properties influence the quality of a matching solution's result.
For this, we evaluate the matching results on datasets they were trained on and others they were not trained on.
Then we evaluate whether correlations between the performance of the data matching solutions and certain dataset characteristics exist.
To do so, we use matching solutions of participants of the \href{https://dbgroup.ing.unimore.it/sigmod21contest/index.shtml}{ACM SIGMOD programming contest 2021}.
The contest consists of three datasets ($D_2$, $D_3$, and $D_4$), which each feature a training dataset ($X_2$, $X_3$ and $X_4$) and a test dataset ($Z_2$, $Z_3$ and $Z_4$) where $Z_i = D_i \setminus X_i$.
The contest participants developed a matching solution for each dataset, based on the training data.
The evaluation data was not provided to the contest participants and was used to assess the quality of the submitted matching solutions.
In our experiment, we simulated a real-world scenario with use case datasets which do not contain a ground truth by executing the data matching solutions of the SIGMOD programming contest on one of the datasets that was not used for their development.
As $D_2$ and $D_3$ share the same schema, we executed matching solutions developed for $D_2$ on $D_3$ and vice versa.

\subsection{Dataset Profiling}

First, we profile the given datasets $D_2$ ($X_2$ \& $Z_2$) and $D_3$ ($X_3$ \& $Z_3$) using the following metrics.
\begin{itemize}
    \itemWithTitle{Sparsity (SP)}{Sparsity is described by Primpeli and Bizer~\cite{Primpeli2020ProfilingMatchingTasks} as the relationship of missing attribute values to all attribute values of the relevant attributes.
    Missing attribute values are challenging for matching solutions and might cause errors~\cite{Petrovski2016SparseData}.}
  
    \itemWithTitle{Textuality (TX)}{Textuality is the average amount of words in attribute values~\cite{Primpeli2020ProfilingMatchingTasks}.
    Non-atomic attributes can complicate the matching task: The matching solution needs to handle long values by tokenizing them during preprocessing or by applying specific similarity measures.}
  
    \itemWithTitle{Tuple count (TC)}{Draisbach and Naumann showed that dataset size has influence on the optimal similarity threshold~\cite{Draisbach2013ChoosingThresholds}.
    Thus, using a benchmark dataset with similar size compared to the use case dataset is preferable. 
    Due to the laborious annotation process, existing benchmark datasets are usally small.
    Draisbach and Naumann further described that, for small datasets, an interval exists for choosing the best similarity threshold. With an increasing number of records, the interval becomes more narrow. Therefore, this fact should be taken into account when selecting a similarity threshold based on a benchmark dataset.}
    
    \itemWithTitle{Positive ratio (PR)}{The positive ratio describes the relationship of the number of true duplicate pairs compared to the number of all pairs. If the number of duplicates (or at least an estimation) is known, the matching solution could use this ratio to dynamically adapt its matching decisions.}
    
    \itemWithTitle{Vocabulary similarity (VS)}{Vocabulary similarity quantifies the similarity of the vocabularies of two datasets.
    Similar vocabularies might cause similar behavior of the matching solution.
    We calculate the vocabulary similarity using the Jaccard coefficient: Let $D_1$, $D_2$ be datasets and let $vocab(D_i)$ be the vocabulary-set of $D_i$, tokenized by spaces.
    \[
    VS(D_1, D_2) = \frac{|vocab(D_1) \cap vocab(D_2)|}{|vocab(D_1) \cup vocab(D_2)|}
    \]
    }
\end{itemize}

Table~\ref{Appendix tab: Profiling the datasets of the ACM SIGMOD programming contest} shows that the datasets have different characteristics and therefore represent different challenges for a matching solution. Both datasets represent the same domain -- notebook specifics. $D_3$ is a lot sparser than $D_2$.
Both have a high textuality, but $D_2$ has a much higher textuality (in average $25.84$).
In the following, we point out the key indicators of the datasets:

\begin{itemize}
    \itemWithTitle{Notebook ($D_2$)} The training dataset as well as the test dataset have a high textuality of $27.99$ and $23.69$.
    A possible matching solution needs to deal with this by, for example, tokenizing the attribute values.
    The vocabulary of $X_2$ and $Z_2$ partly overlaps. Approximately every third token exists in both datasets.
   
    \itemWithTitle{Notebook large ($D_3$)} $X_3$ as well as $Z_3$ are sparse -- approximately every second attribute is missing ($50.1\%$ \& $42.6\%$).
    Furthermore, the datasets include much textual data with an average attribute length of 15 words.
    The training and test dataset of D3 differs in its characteristics.
    $2.2\%$ of all possible pairs in the training dataset $X_3$ are duplicates.
    In contrast, $12.1\%$ of all pairs in the test dataset $Z_3$ are duplicates.
    This gap might influence the quality performance of a matching solution.
\end{itemize}

\begin{table}[ht]
\centering
\begin{tabular}{ m{1.1cm} | m{0.8cm} | m{0.8cm} | m{0.8cm} | m{0.8cm}}
&
    \multicolumn{2}{c|}{$D_2$ dataset}
&
    \multicolumn{2}{c}{$D_3$ dataset}
\\
    \multicolumn{1}{c|}{Dataset}
&
   \multicolumn{1}{c|}{Train ($X_2$)}
&
    \multicolumn{1}{c|}{Test ($Z_2$)}
&
   \multicolumn{1}{c|}{Train ($X_3$)}
&
    \multicolumn{1}{c}{Test ($Z_3$)}
\\\hline

    SP
&
    \multicolumn{1}{r|}{11.1\%}
&
    \multicolumn{1}{r|}{19.72\%}
&
    \multicolumn{1}{r|}{50.1\%}
&
    \multicolumn{1}{r}{42.6\%}
\\\hline

    TX
&
    \multicolumn{1}{r|}{27.99}
&
    \multicolumn{1}{r|}{23.69}
&
    \multicolumn{1}{r|}{15.53}
&
    \multicolumn{1}{r}{15.35}
\\\hline
    TC
&
    \multicolumn{1}{r|}{58'653}    
&
    \multicolumn{1}{r|}{18'915}
&
    \multicolumn{1}{r|}{56'616}
&
    \multicolumn{1}{r}{35'778}
   \\\hline
    PR
&
    \multicolumn{1}{r|}{2.2\%}
&
    \multicolumn{1}{r|}{3.6\%}
&
    \multicolumn{1}{r|}{2.2\%}
&
    \multicolumn{1}{r}{12.1\%}
    \\\hline
    VS
&
    \multicolumn{2}{c|}{59.0\%}
&
    \multicolumn{2}{c}{37.7\%}

\end{tabular}
\captionWithDescription{Profiling the datasets of the ACM SIGMOD programming contest}{}
\label{Appendix tab: Profiling the datasets of the ACM SIGMOD programming contest}
\end{table}

\subsection{Influence of Dataset Characteristics on Matching Solutions}
In the following, we analyze the influence of specific dataset characteristics on the average quality of matching solutions.
We executed three matching solutions of the SIGMOD contest on $X_2$, $X_3$, $Z_2$ and $Z_3$ and loaded the results into \Snowman. 
\Snowman supports this evaluation by providing an easy way to determine the quality metrics of experiments and to choose the best similarity threshold for each matching solution.
In the next part, we always refer to the average quality metrics over these solutions. 

As shown in Table~\ref{Appendix tab: Average quality metrics of matching solution runs of the ACM SIGMOD programming contest}, it becomes clear that matching solutions generally perform better on the datasets on which they have been developed than on new data, as expected.
Taking a look at the average \fMeasure, one can observe that the matching solution developed for $D_3$ performs a lot better on $D_2$ (average \fMeasure $80.5\%$) than the matching solution developed for $D_2$ does on $D_3$ (average \fMeasure $41.4\%$).
As outlined above, the $D_3$ dataset is a sparse dataset with in average $46.4\%$ missing attribute values.
The matching solutions trained on a sparse dataset performed better on a non-sparse dataset than the matching solutions that were developed on a non-sparse dataset applied to a sparse dataset.
Therefore, one reason for the poor performance might be the missing data.
Thus, when selecting a benchmark dataset, the sparsity should be similar to the real-world dataset or lower.

Additionally, one can observe a gap between the average quality metrics of the test- and training dataset of $D_3$.
The test dataset $Z_3$ has an average \fMeasure of $35.7\%$ and the training dataset $X_3$ has an average \fMeasure of $47.0\%$, resulting in a difference of $\Delta$\fMeasure$=11.3\%$.
The difference in the average \fMeasure is much smaller for $X_2$ and $Z_2$: $X_2$ has an average \fMeasure of $81.3\%$ and $Z_2$ has an average \fMeasure of $79.6\%$.
This results in a difference of $\Delta$\fMeasure$=1.7\%$.
One reason for this might be the lower vocabulary similarity of $X_3$ and $Z_3$.
The vocabularies of the datasets $X_2$ and $Z_2$ are a lot more similar ($59.0\%$) than the vocabularies of $X_3$ and $Z_3$ ($37.7\%$).
It might be easier for the classification model, developed on $X_2$, to classify the candidate pairs correctly, as its vocabulary is more similar to the vocabulary used during training.
In summary, vocabulary similarity might be a relevant indicator for the suitability of benchmark datasets.

\begin{table}[ht]
\centering
\begin{tabular}{ m{1.0cm} | m{1.4cm} | m{0.8cm} | m{0.8cm} | m{0.8cm} | m{0.8cm}}
    \multicolumn{1}{c|}{\multirow{2}{1.2cm}{Matching solution}}
&
    \multicolumn{1}{c|}{\multirow{2}{*}{Metric}}
&
    \multicolumn{2}{c|}{$D_2$ dataset}
&
    \multicolumn{2}{c}{$D_3$ dataset}
\\
&
&
   \multicolumn{1}{c|}{Train}
&
    \multicolumn{1}{c|}{Test}
&
   \multicolumn{1}{c|}{Train}
&
    \multicolumn{1}{c}{Test}
\\\hline
\multicolumn{1}{c|}{\multirow{3}{1.2cm}{developed on $X_2$}}
&
    Precision
&
    \multicolumn{1}{r|}{100\%}
&
    \multicolumn{1}{r|}{97.7\%}
&
    \multicolumn{1}{r|}{46.9\%}
&
    \multicolumn{1}{r}{90.1\%}
\\

&
    Recall
&
    \multicolumn{1}{r|}{99.6\%}
&
    \multicolumn{1}{r|}{97.0\%}
&
    \multicolumn{1}{r|}{56.2\%}
&
    \multicolumn{1}{r}{43.2\%}
\\
&
    \fMeasure
&
    \multicolumn{1}{r|}{99.8\%}
&
    \multicolumn{1}{r|}{97.4\%}
&
    \multicolumn{1}{r|}{35.7\%}
&
    \multicolumn{1}{r}{47.0\%}
\\\hline
\multicolumn{1}{c|}{\multirow{3}{1.2cm}{developed on $X_3$}}
&
    Precision
&
    \multicolumn{1}{r|}{76.3\%}
&
    \multicolumn{1}{r|}{68.5\%}
&
    \multicolumn{1}{r|}{69.7\%}
&
    \multicolumn{1}{r}{98.6\%}
   \\
&
    Recall
&
    \multicolumn{1}{r|}{89.5\%}
&
    \multicolumn{1}{r|}{95.0\%}
&
    \multicolumn{1}{r|}{97.2\%}
&
    \multicolumn{1}{r}{97.5\%}
    \\
&
    \fMeasure
&
    \multicolumn{1}{r|}{81.3\%}
&
    \multicolumn{1}{r|}{79.6\%}
&
    \multicolumn{1}{r|}{76.5\%}
&
    \multicolumn{1}{r}{98.2\%}
\end{tabular}

\captionWithDescription{Average quality metrics of matching solutions of the ACM SIGMOD programming contest}{
The first row contains average results of matching solutions that were developed on dataset $X_2$. The second row contains average results of matching solutions that were developed on dataset $X_3$.}
\label{Appendix tab: Average quality metrics of matching solution runs of the ACM SIGMOD programming contest}
\end{table}

\subsection{Conclusion}
The characteristics of a dataset have a traceable influence on the performance of a matching solution.
Thus, when choosing a benchmark dataset, the suitability of the properties of the benchmark dataset should be considered.
Concretely, a benchmark dataset should present a matching solution with similar challenges as the use case dataset under consideration. If the datasets differ too much in their properties, the matching results are not representative.

A next step is to confirm the findings on more datasets and matching solutions.
Finally, a similarity measure between datasets for the selection of a suitable benchmark dataset based on the examined dataset properties needs to be developed.

\section{Efficiently Calculating Metric/Metric Diagrams}
\label{Appendix ssec: Constructing Dynamic Intersections for High Performance Precision/Recall Diagrams}

Many matching solutions use a similarity threshold to distinguish duplicates from non-duplicates:
A pair is matched if and only if its similarity score is greater than or equal to the threshold (see Section~\ref{ssec: Formal Matching Process}).
Therefore, the similarity threshold has a large impact on matching quality.
To assist users in finding good similarity thresholds, \frost includes metric/metric diagrams, such as precision/recall or \fMeasure/similarity diagrams (see Section~\ref{sssec: Metric-Metric Diagrams}).
The data points of these diagrams correspond to pair-based quality metrics of the matching solution based on different similarity thresholds.
Thus, they give the user an overview about the general performance of the matching solution and about which similarity thresholds and similarity functions work well.

As pair-based metrics can be calculated in constant time from a confusion matrix, an algorithm for calculating metric/metric diagrams can simply output a list of confusion matrices corresponding to the requested similarity thresholds.
Thus, a na\"ive approach to calculating metric/metric diagrams is to go through the list of matches and track all sets of pairs in the confusion matrix.
A problem with this approach is that to align with \snowman's concept of experiments, the matches at each step need to be transitively closed.
Thus, the total number of pairs is quadratic, which means that the algorithm has quadratic runtime in the size of the dataset.
A slightly more advanced (but still na\"ive) approach could utilize the fact that the confusion matrix of an experiment and ground truth annotations can be calculated in linear runtime by representing experiment and ground truth as clusterings and calculating the intersection between those clusterings.
It could then calculate the experiment clustering, intersection, and confusion matrix newly for every requested similarity threshold.
But again, while drawing metric/metric diagrams with this approach is no problem for small datasets, the runtime increases rapidly as dataset size increases (see Table~\ref{Tab: Runtime of Back-end Algorithms} on page~\pageref{Tab: Runtime of Back-end Algorithms}).
This comes from the fact that the runtime is linear in the \emph{product} of the number of requested thresholds and dataset length in both worst and best case.
This makes this algorithm unsuitable for large datasets.
However, this functionality can be optimized by reusing intermediate results.
While this is simple for the experiment clustering, updating the intersection clustering is a challenging task because matches do not necessarily have a direct impact on the intersection clustering, but still can cause side effects later.
An example is depicted in Figure~\ref{Appendix fig: Intersecting two Clusterings}: Consider the ground truth clustering $\{\{a,b\}, \{c\}\}$ and the matches $\{b,c\}$ and $\{a,c\}$.
The intersection does not change after merging $\{b,c\}$ because $b$ and $c$ are in different ground truth clusters.
The same holds true for $\{a,c\}$, but because $b$ and $c$ already have been merged together before, the intersection now contains the cluster $\{a,b\}$.

\begin{figure}[t]
\centering
\includegraphics[width=4.5cm]{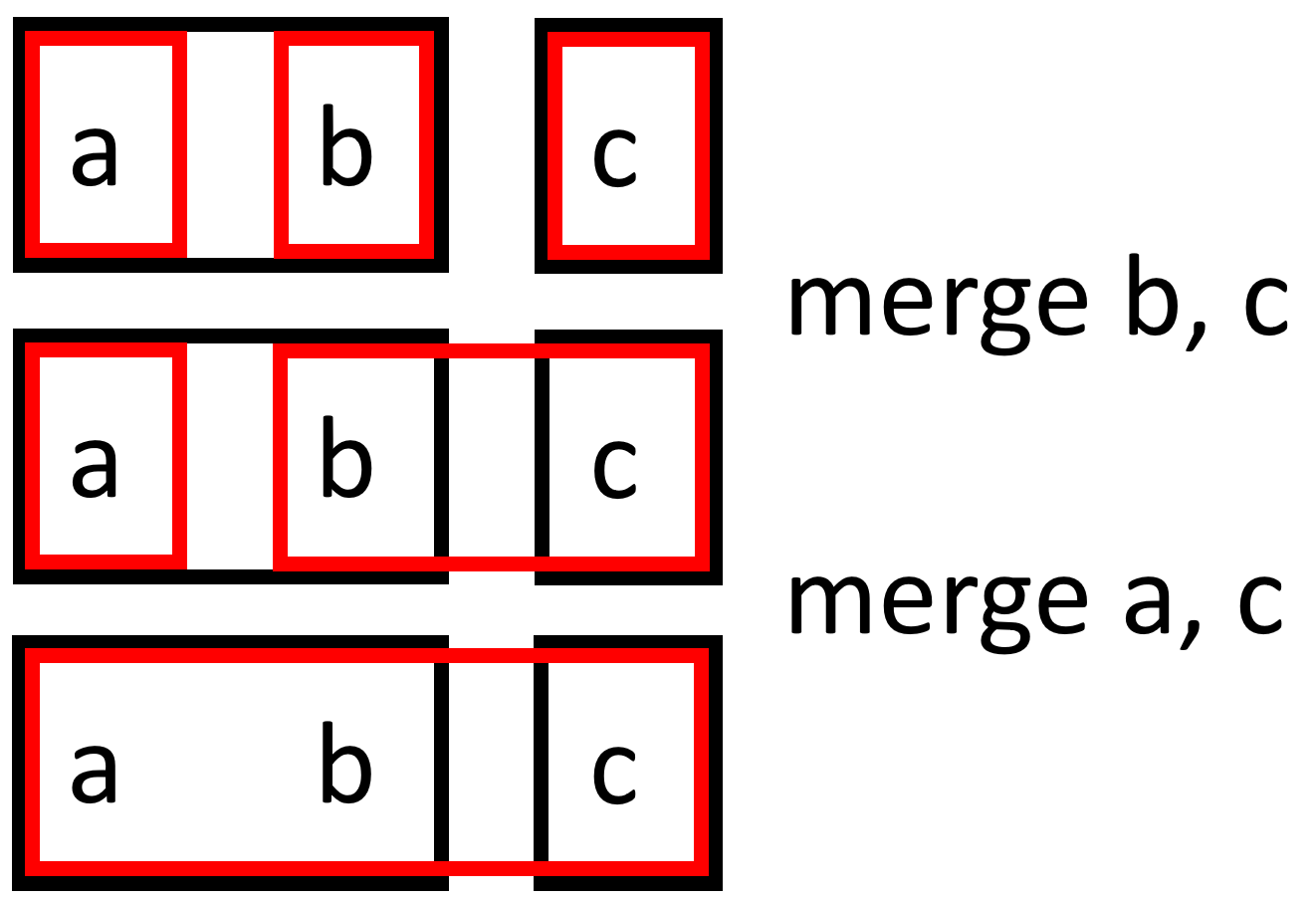}
\captionWithDescription{Intersecting two Clusterings}{ground truth clustering: black, experiment clustering: red, intersection: black and red}
\label{Appendix fig: Intersecting two Clusterings}
\end{figure}

Here, we discuss a data structure which efficiently solves this issue by explaining the algorithm used by \snowman to calculate a sequence of confusion matrices belonging to different similarity thresholds.
In short, it constructs a clustering of the experiment and a clustering of the intersection between experiment and ground truth.
For each requested similarity threshold, it reuses the clustering and intersection from the last requested threshold.
Meanwhile, it tracks the count of pairs in experiment and intersection clustering, which allows it to efficiently calculate the confusion matrix for each requested threshold.
Figure~\ref{Appendix fig: Exemplary Construction of the Intersection} illustrates the process of creating experiment and intersection clustering.
The example is explained step by step throughout the section.

\subsection{Input and Output}

For clarity, we capitalize and pluralize collections (\textit{Numbers}) and lowercase and singularize everything else (\textit{number}).
Clusterings are denoted as $C_\mathit{<identifier>}$.
We denote the number of elements of a collection with $|\cdot|$.
The algorithm receives the following parameters as input:

\begin{itemize}
    \item $D$: the dataset that the matching solution was executed on
    
    \item \textit{Matches}: array of matches predicted by the matching solution.
        Every entry of the array contains a similarity score and the two merged records.
    
    \item $C_\mathit{truth}$: ground truth duplicate clustering of $D$, assigning a cluster ID to every record
    
    \item \textit{s}: the number of similarity scores at which metrics should be sampled.
        For simplicity, we assume that $|\mathit{Matches}|$ is divisible by $s - 1$.
\end{itemize}

As output, the algorithm produces a list \textit{Matrices} containing \textit{s} confusion matrices.
The na\"ive way to sample \textit{Matrices} is by increasing the similarity threshold by a constant amount every time.
But this has the flaw that imbalances in the distribution of thresholds can lead to data points which have a lot of matches between them and data points which have no matches between them.
To combat this, the matrices are created by increasing the number of matches between two matrices by a constant amount.
For example, if the similarity scores $0$, $1$, $3$, and $6$ are present and $3$ matrices should be calculated, \textit{Matrices} contains matrices for the similarity thresholds \textit{infinity} (no matches), $3$, and $0$.
By this, between the thresholds \textit{infinity} and $3$ lie the matches with the scores $6$ and $3$ and between the thresholds $3$ and $0$ lie the matches with the scores $1$ and $0$.
Formally, the $i$-th entry of \textit{Matrices} contains the confusion matrix of the matching solution based on the similarity threshold of the entry in \textit{Matches} that has the $(i \cdot \frac{|\mathit{Matches}|}{\mathit{s} - 1})$ biggest similarity score.
If $i$ is zero, \textit{Matches} contains the confusion matrix of the matching solution with similarity threshold infinity.

To efficiently update the intersection clustering, we need a mechanism to know which clusters of the experiment clustering were merged between the current and the last threshold.
For brevity, we assume that the union-find data structure~\cite{Tarjan1972UnionFind} provides a mechanism for this by supporting the operation \textit{trackedUnion}.
\textit{trackedUnion} works like a batched \textit{union}, which outputs a list of which clusters were merged.
The operation takes a list \textit{Pairs} of record pairs as input.
It calls union for the two clusters of every pair generating a new cluster ID for the resulting cluster.
Then, it outputs a list \textit{Merges}, which contains an entry for every newly created cluster that has not already been merged with another cluster.
Every entry stores the ID of the newly created cluster \textit{target} and a list \textit{Sources} of cluster IDs from before the update, which are now part of \textit{target}.

As an example, let the clustering contain the three clusters $\{\{\mathit{a}\}, \{\mathit{b}\}, \{\mathit{c, d}\}\}$ and \textit{Pairs} contain the record pairs $\{\mathit{a},\mathit{b}\}$ and $\{\mathit{b},\mathit{c}\}$. 
After executing \textit{trackedUnion}, the clustering would contain the cluster $\{\{\mathit{a, b, c, d}\}\}$ and \textit{trackedUnion} would return \textit{Merges} that contain exactly one entry.
The entry would be composed of the IDs of the clusters $\{\mathit{a}\}$, $\{\mathit{b}\}$, and $\{\mathit{c, d}\}$ as \textit{Sources} and the ID of the cluster $\{\mathit{a},\mathit{b},\mathit{c}, \mathit{d}\}$ as \textit{target}.
An example of \textit{trackedUnion} for multiple consecutive invocations can be seen in Figure~\ref{Appendix fig: Exemplary Construction of the Intersection} (\textit{trackedUnion} is executed on $C_\mathit{exp}$).

We also need a method to calculate the number of pairs in experiment clustering and intersection clustering efficiently.
For this, we assume that the union-find data structure can track the number of pairs in each of its clusters and overall.

\subsection{Algorithm Description}

Algorithm~\ref{Appendix alg: Compute Confusion Matrices} depicts the algorithm used by \snowman to calculate the list of confusion matrices described above.
First, it creates the initial state of the experiment clustering $C_\mathit{exp}$ using a union-find data structure with a cluster of size one for every record of dataset~$D$.
Then, it initializes the intersection clustering $C_\mathit{intersect}$ of experiment clustering $C_\mathit{exp}$ and ground truth clustering $C_\mathit{truth}$ as described in Section~\ref{Appendix ssec: Dynamically Constructing the Intersection}.
Next, the algorithm sets the output list \textit{Matrices} to an empty list and adds the confusion matrix for the initial state of $C_\mathit{exp}$, $C_\mathit{truth}$, and $C_\mathit{intersect}$.
Now, it sorts \textit{Matches} by similarity score in descending order and splits the result into $\mathit{s} - 1$ ranges.

For each range, first it updates the experiment clustering $C_\mathit{exp}$ with the \textit{trackedUnion} function, saving its result into the list \textit{Merges}.
Then, it updates the dynamic intersection $C_\mathit{intersect}$ with the help of \textit{Merges} as described in Section~\ref{Appendix ssec: Dynamically Constructing the Intersection}.
Finally, the algorithm adds the confusion matrix for the current state of $C_\mathit{exp}$, $C_\mathit{truth}$, and $C_\mathit{intersect}$ to the output list \textit{Matrices}.
After the loop terminates, \textit{Matrices} contains $s$ confusion matrices.

\begin{figure*}[t]
\centering
\begin{tabular}{ m{0.6cm} | m{0.8cm} | m{2cm} | m{2.2cm} | m{2cm} | m{8cm} }
    Step
&
    Merge
&
    $C_\mathit{exp}$
&
    $\mathit{trackedUnion}(...)$
&
    $C_\mathit{intersect}$
&
    Confusion Matrix (only the numbers are stored)
\\\hline

    0
&
&
    \begin{tabular}{ m{0.3cm} m{1.0cm} } 
        \textit{e0} & $a$
    \\\hdashline
        \textit{e1} & $b$ 
    \\\hdashline
        \textit{e2} & $c$
    \\\hdashline
        \textit{e3} & $d$
    \end{tabular}
&
&
    \begin{tabular}{ m{0.3cm} m{1.0cm} } 
        \textit{e0} & \textit{g0}: $a$ 
    \\\hdashline
        \textit{e1} & \textit{g0}: $b$
    \\\hdashline
        \textit{e2} & \textit{g1}: $c$
    \\\hdashline
        \textit{e3} & \textit{g1}: $d$
    \end{tabular}
& 
    \begin{tabular}{ m{2.8cm} | m{4.5cm} }  
        TP: $0$ $(\{\})$ & FP: $0$ $(\{\})$ 
    \\\hline
        FN: $2$ $(\{\{a,b\}, \{c,d\}\})$ & TN: $4$ $(\{\{a,c\}, \{a,d\}, \{b,c\}, \{b,d\}\})$
    \\
    \end{tabular}
\\\hline

    1   
&
     $\{\mathit{a}, \mathit{c}\}$
&
    \begin{tabular}{ m{0.3cm} m{1.0cm} } 
        \textit{e1} & $b$
    \\\hdashline
        \textit{e3} & $d$
    \\\hdashline
        \textit{e4} & $a$, $c$  \\
    \end{tabular} 
&
    \begin{tabular}{ m{0.4cm} m{0.9cm} } 
        \textit{src} & \textit{e0}, \textit{e2}
        \\\hdashline
        \textit{tgt} & \textit{e4}
    \end{tabular}
&
    \begin{tabular}{ m{0.3cm} m{1.0cm} } 
        \textit{e1} & \textit{g0}: $b$ 
    \\\hdashline
        \textit{e3} & \textit{g1}: $d$
    \\\hdashline
        \textit{e4} & \textit{g0}: $a$ \newline
               \textit{g1}: $c$
    \end{tabular}
& 
    \begin{tabular}{ m{2.8cm} | m{4.5cm} }  
        TP: $0$ $(\{\})$ & FP: $1$ $(\{\{a,c\}\})$ 
    \\\hline
        FN: $2$ $(\{\{a,b\}, \{c,d\}\})$ & TN: $3$ $(\{\{a,d\}, \{b,c\}, \{b,d\}\})$
    \\
    \end{tabular}
\\\hline

    2 
&
     $\{\mathit{b}, \mathit{d}\}$
&
    \begin{tabular}{ m{0.3cm} m{1.0cm} } 
        \textit{e4} & $a$, $c$
    \\\hdashline
        \textit{e5} & $b$, $d$
    \end{tabular} 
&
    \begin{tabular}{ m{0.4cm} m{0.9cm} } 
        \textit{src} & \textit{e1}, \textit{e3}
        \\\hdashline
        \textit{tgt} & \textit{e5}
    \end{tabular}
&
    \begin{tabular}{ m{0.3cm} m{1.0cm} } 
        \textit{e4} & \textit{g0}: $a$ \newline
               \textit{g1}: $c$
    \\\hdashline
        \textit{e5} & \textit{g0}: $b$ \newline
               \textit{g1}: $d$
    \end{tabular}
& 
    \begin{tabular}{ m{2.8cm} | m{4.5cm} }  
        TP: $0$ $(\{\})$ & FP: $2$ $(\{\{a,c\}, \{b,d\}\})$ 
    \\\hline
        FN: $2$ $(\{\{a,b\}, \{c,d\}\})$ & TN: $2$ $(\{\{a,d\}, \{b,c\}\})$
    \\
    \end{tabular}
\\\hline

    3
&
     $\{\mathit{a}, \mathit{b}\}$
&
    \begin{tabular}{ m{0.3cm} m{1.0cm} } 
        \textit{e6} & $a$, $b$, $c$, $d$
    \end{tabular}
&
    \begin{tabular}{ m{0.4cm} m{0.9cm} } 
        \textit{src} & \textit{e4}, \textit{e5}
        \\\hdashline
        \textit{tgt} & \textit{e6}
    \end{tabular}
&
    \begin{tabular}{ m{0.3cm} m{1.0cm} } 
        \textit{e6} & \textit{g0}: $a$, $b$ \newline
               \textit{g1}: $c$, $d$
    \end{tabular}
& 
    \begin{tabular}{ m{2.8cm} | m{4.5cm} }  
        TP: $2$ $(\{\{a,b\}, \{c,d\}\})$ & FP: $4$ $(\{\{a,c\}, \{a,d\}, \{b,c\}, \{b,d\}\})$ 
    \\\hline
        FN: $0$ $(\{\})$ & TN: $0$ $(\{\})$
    \\
    \end{tabular}
\\
\end{tabular}
\captionWithDescription{Exemplary run of the algorithm}{
Example for dataset $\{\mathit{a}, \mathit{b}, \mathit{c}, \mathit{d}\}$, ground truth clustering \textit{g0}: $\{\mathit{a}, \mathit{b}\}$, \textit{g1}: $\{\mathit{c}, \mathit{d}\}$,
and detected matches  $\{\mathit{a}, \mathit{c}\}$, $\{\mathit{b}, \mathit{d}\}$, $\{\mathit{a}, \mathit{b}\}$.
The example run is described in detail in Section~\ref{Appendix ssec: Exemplary Run of the Algorithm}.}
\label{Appendix fig: Exemplary Construction of the Intersection}
\end{figure*}

\subsection{Dynamically Constructing the Intersection}
\label{Appendix ssec: Dynamically Constructing the Intersection}

The intersection clustering $C_\mathit{intersect}$ is stored as a pair of two variables: 
A union-find data structure for calculating the number of pairs and a map for speeding up the construction of the intersection.
Each intersection cluster in the union-find data structure is uniquely identified by an experiment cluster and a ground truth cluster, and contains all records they have in common.
The map uses this property as follows:
The map contains an entry for every cluster of the experiment clustering $e$ mapping to yet another map from every involved ground truth cluster $g$ to the intersection cluster of $e$ and $g$.
For an example see Figure~\ref{Appendix fig: Exemplary Construction of the Intersection} column $C_\mathit{intersect}$.

Initially, the experiment clustering has a cluster of size one for every record in $D$ (see Figure~\ref{Appendix fig: Exemplary Construction of the Intersection} row one column $C_\mathit{exp}$).
Therefore, the initial state of the $C_\mathit{intersect}$ union-find data structure contains a cluster of size one for every record in $D$.
Hence, the initial state of the $C_\mathit{intersect}$ map maps from every initial experiment cluster $\{\mathit{r}\}$ to a map containing exactly one \textit{key} $\rightarrow$ \textit{value} pair, namely \textit{ground truth cluster of r} $\rightarrow$ \textit{intersection cluster of r} (see Figure~\ref{Appendix fig: Exemplary Construction of the Intersection} row one column $C_\mathit{intersect}$).

To update $C_\mathit{intersect}$ with a list of merged clusters as returned by \textit{trackedUnion}, Algorithm~\ref{Appendix alg: Update Dynamic Intersection} iterates over each pair of \textit{Sources} clusters and \textit{target} cluster.
First, it aggregates all intersection clusters belonging to the \textit{Sources} clusters into a list of involved intersection clusters.
Now, it groups the list by ground truth cluster into a map from ground truth cluster to intersection clusters.
Then, it loops over all \textit{ground truth cluster} $\rightarrow$ \textit{intersection clusters} pairs and merges the intersection clusters into a new intersection cluster (the merge happens in the union-find data structure).
Finally, the algorithm stores the newly created intersection clusters into a map from ground truth cluster to new intersection cluster (also contains old clusters if they were in a list of size one) and saves it into the map of $C_\mathit{intersect}$ at the position of the \textit{target} cluster.

For an example update process, see the listing below.
The example shows the process of merging $a$ and $b$ from Figure~\ref{Appendix fig: Exemplary Construction of the Intersection} (there is only one pair of \textit{Sources} clusters and \textit{target} cluster):

\begin{itemize}
    \label{Appendix listing: Example Dynamic Intersection Update}
    \itemWithTitle{List of involved intersection clusters}{\\
    \textit{e4g0} ($a$), \textit{e4g1} ($c$), \textit{e5g0} ($b$), \textit{e5g1} ($d$)}
    
    \itemWithTitle{Map from ground truth cluster to intersection clusters}{
    \textit{g0}: \textit{e4g0}, \textit{e5g0};
    \textit{g1}: \textit{e4g1}, \textit{e5g1}}
    
    \itemWithTitle{Merge the intersection clusters}{\\
    $\mathit{e4g0}, \mathit{e5g0} \rightarrow \mathit{e6g0}$;
    $\mathit{e4g1}, \mathit{e5g1} \rightarrow \mathit{e6g1}$}
    
    \itemWithTitle{Map from ground truth cluster to new intersection cluster}{
    \textit{g0}: \textit{e6g0};
    \textit{g1}: \textit{e6g1}}
\end{itemize}

\begin{algorithm}[t]
    \caption{Compute Confusion Matrices}
    \label{Appendix alg: Compute Confusion Matrices}
    \begin{algorithmic}[1]
        \STATE $C_\mathit{exp} \leftarrow$ new \textit{UnionFind} with $|\mathit{D}|$ clusters of size $1$
        
        \STATE $C_\mathit{intersect} \leftarrow$ initial intersection clustering (see Section~\ref{Appendix ssec: Dynamically Constructing the Intersection})
        
        \STATE $\mathit{Matrices} \leftarrow$ new list
        
        \STATE $\mathit{Matrices}.\mathit{append}(\mathit{getConfusionMatrix}(C_\mathit{exp}, C_\mathit{truth}, C_\mathit{intersect}))$
        
        \STATE sort \textit{Matches} by similarity score in descending order
        
        \FOR{$i$ from $1$ to $s - 1$}
            \STATE \textit{start} $\leftarrow$ $(\mathit{i}-1) \cdot \frac{|\mathit{Matches}|}{\mathit{s} - 1}$
        
            \STATE \textit{stop} $\leftarrow \mathit{i} \cdot \frac{|\mathit{Matches}|}{\mathit{s} - 1}$
        
            \STATE \textit{Pairs} $\leftarrow \mathit{Matches}[\mathit{start}:\mathit{stop} - 1]$
        
            \STATE \textit{Merges} $\leftarrow C_\mathit{exp}.\mathit{trackedUnion}(\mathit{Pairs})$
        
            \STATE update $C_\mathit{intersect}$ with \textit{Merges} as described in Algorithm~\ref{Appendix alg: Update Dynamic Intersection}
        
            \STATE $\mathit{Matrices}.\mathit{append}(\mathit{getConfusionMatrix}(C_\mathit{exp}, C_\mathit{truth}, C_\mathit{intersect}))$
        \ENDFOR
        
        \RETURN \textit{Matrices}
    \end{algorithmic}
\end{algorithm}

\begin{algorithm}[t]
    \caption{Update Dynamic Intersection}
    \label{Appendix alg: Update Dynamic Intersection}
    \begin{algorithmic}[1]
        \FOR{(\textit{Sources}, \textit{target}) in \textit{Merges}}
            \STATE \textit{intersectionClusters} $\leftarrow$ empty list
            
            \FOR{\textit{source} in \textit{Sources}}
                \STATE $\mathit{intersectionClusters}.\mathit{appendAll}(C_\mathit{intersect}[\mathit{source}].\mathit{values}())$
            \ENDFOR
            
        \STATE group \textit{intersectionClusters} by ground truth cluster
        
        \STATE $C_\mathit{intersect}[\mathit{target}] \leftarrow$ empty map
            
        \FOR{(\textit{truthCluster},\\
        \:\:\:\:\:\:\:\: \textit{intersectionClusterGroup}) in \textit{intersectionClusters}}
            \STATE \textit{newCluster} $\leftarrow C_\mathit{intersect}.\mathit{unionAll}(\mathit{intersectionClusterGroup})$
            
            \STATE $C_\mathit{intersect}[\mathit{target}][\mathit{truthCluster}] \leftarrow$ \textit{newCluster}
        \ENDFOR
        \ENDFOR
    \end{algorithmic}
\end{algorithm}

\subsection{Exemplary Run of the Algorithm}
\label{Appendix ssec: Exemplary Run of the Algorithm}

Let the deduplicated dataset $D$ contain the four entries $a$, $b$, $c$, and $d$.
Let the ground truth clustering $C_\mathit{truth}$ contain the two clusters $\mathit{g0}: \{\mathit{a}, \mathit{b}\}$ and $\mathit{g1}: \{\mathit{c}, \mathit{d}\}$ (\textit{g0} and \textit{g1} are the IDs of the clusters).
Let the detected matches \textit{Matches} of the matching solution be $\{a, c\}$, $\{b, d\}$, and $\{a, b\}$.
Let \textit{s} be $4$.
That means a confusion matrix should be calculated after merging each pair (and before merging the first).

Figure~\ref{Appendix fig: Exemplary Construction of the Intersection} shows the experiment clustering $C_\mathit{exp}$, the result of \textit{trackedUnion}, the intersection clustering $C_\mathit{intersect}$, and the resulting confusion matrix after merging each pair (and before merging the first).
Hence, the output of the algorithm contains a sequence of all depicted confusion matrices.
For example, the second row (step $1$) contains the state after $a$ and $c$ have been merged.
Concretely, the experiment clustering contains the three clusters  $\{\mathit{b}\}$, $\{\mathit{d}\}$, and $\{\mathit{a}, \mathit{c}\}$ with the cluster IDs \textit{e1}, \textit{e3}, and \textit{e4}.
Because the records $a$ and $c$ have been merged, their respective clusters $e0$ and $e2$ from the initial experiment clustering are listed as the \textit{source} clusters of \textit{trackedUnion}.
Because the cluster which contains $a$ and $c$ now has the id \textit{e4}, the \textit{target} cluster is \textit{e4}.
The intersection clustering contains the four clusters  $\{\mathit{b}\}$, $\{\mathit{d}\}$, $\{\mathit{a}\}$, and $\{\mathit{c}\}$.
In the table, every intersection cluster is shown to the right of the experiment cluster ID and ground truth cluster ID whose intersection it represents.
For example, the intersection cluster $\{\mathit{a}\}$ represents the intersection of the experiment cluster \textit{e4} and the ground truth cluster \textit{g0}.
Therefore, it is shown to the right of \textit{e4} and \textit{g0}.
Note that this exactly depicts the map data structure of the intersection clustering.
The confusion matrix has no true positives, one false positive, two false negatives, and three true negatives.
Observe that the number of true positives equals the number of pairs in $C_\mathit{intersect}$.

\subsection{Conclusion}

In summary, we presented \snowman's optimized algorithm for computing metric/metric diagrams.
For that, we presented an operation that dynamically constructs the intersection of two clusterings.
A runtime analysis of the algorithm shows that the worst-case runtime of the algorithm is in $O(|\mathit{D}| + |\mathit{Matches}| \cdot (\mathit{s} + log(|\mathit{Matches}|))$.
When the ground truth has fewer than $|\mathit{D}|$ clusters, the complexity is even better (if it has $O(1)$ clusters, the runtime is in $O(|\mathit{D}| + |\mathit{Matches}| \cdot log(|\mathit{Matches}|)$).
Additionally, the algorithm runs the faster, the more similar ground truth and experiment clusterings are.
Note, that \snowman optimizes experiments when they are uploaded, which means that users of \snowman will experience a runtime between $O(|\mathit{D}|)$ and $O(|\mathit{D}| + |\mathit{Matches}| \cdot \mathit{s})$, and \textit{Matches} only contains pairs that cannot be inferred by transitively closing.
Table~\ref{Tab: Runtime of Back-end Algorithms} confirms that the algorithm is considerably faster compared to the na\"ive approach.

An interesting extension to metric/metric diagrams is a timeline feature in which new true positives and false positives between two similarity thresholds are shown (note that \frost's and \snowman's set-based comparisons already allow this).
While the concepts of the above algorithm optimize this use case to some extent, the dynamic intersection and union find data structure lack the functionality to ``revert'' merges: whenever the user selects a similarity threshold range starting before the end of the previous range, $O(|\mathit{D}|)$ time is necessary to reset the clusterings.
This makes interactively exploring the timeline slow, especially for large datasets.
Therefore, a useful next step is to develop an algorithm for efficiently reverting merges in the dynamic intersection and union find data structure.

}

\end{document}